\documentclass[10pt,dvipsnames,x11names,table]{article}
\usepackage[T1]{fontenc}
\usepackage{libertine}
\usepackage[libertine]{newtxmath}
\usepackage[scale=0.875]{sourcecodepro}

\usepackage{algorithm,algorithmic}
\usepackage{booktabs}
\usepackage[font={rm}]{caption}
\usepackage{doi}
\usepackage{enumerate}
\usepackage{enumitem}
\usepackage{float,multicol}
\usepackage[margin=1in]{geometry}
\usepackage[htt]{hyphenat}
\usepackage{hyperref}
\usepackage{graphicx}
\usepackage{listings}
\usepackage{multirow}
\usepackage{newfloat}
\usepackage{subcaption}
\usepackage{svg}
\usepackage{tabto}
\usepackage{textcomp}
\usepackage{threeparttable}
\usepackage{tikz}
\usetikzlibrary{positioning, shapes, arrows.meta, decorations.text}
\usepackage{url}
\usepackage{verbatim}
\usepackage{enumitem}

\DeclareFloatingEnvironment[fileext=diag,placement={!ht},name=Diagram]{diag}

\makeatletter
\renewcommand{\@font@warning}[1]{}
\renewcommand{\@latex@warning}[1]{}
\makeatother

\newfloat{lstfloat}{ht}{lop}
\floatname{lstfloat}{Listing}

\DeclareCaptionSubType{lstfloat}

\definecolor{linkcolor}{RGB}{139,0,0}
\definecolor{citecolor}{RGB}{85,26,139}
\definecolor{linkblue}{RGB}{0,102,153}

\hypersetup{
    breaklinks,
    colorlinks=true,
    linkcolor=linkcolor,
    citecolor=citecolor,
    urlcolor=linkblue,
}

\newcommand{\footnotesmallb}{\fontsize{8.7}{10.4}\selectfont}

\begin{document}

\begin{center}
    \Large\bfseries\sffamily
    A Critical Roadmap to Driver Authentication via CAN Bus: \\ Dataset Review, Introduction of the Kidmose CANid Dataset (KCID), \\ and Proof of Concept
\par\end{center}
\vspace{0.5em}

\begin{center}
  \normalsize\sffamily
  \begin{minipage}[t]{0.33\textwidth}
    \centering
    \textbf{Brooke Elizabeth Kidmose}\\
    Department of Applied Mathematics\\
    and Computer Science\\
    Technical University of Denmark\\
    Kgs. Lyngby, Denmark\\
    \texttt{blam@dtu.dk}\\
  \end{minipage}%
  \hfill
  \begin{minipage}[t]{0.33\textwidth}
    \centering
    \textbf{Andreas Brasen Kidmose}\\
    Department of Electrical and\\
    Photonics Engineering\\
    Technical University of Denmark\\
    Kgs. Lyngby, Denmark\\
    \texttt{abki@dtu.dk}\\
  \end{minipage}%
  \hfill
  \begin{minipage}[t]{0.33\textwidth}
    \centering
    \textbf{Cliff C. Zou}\\
    Department of Computer Science\\
    University of Central Florida\\
    Orlando, Florida, USA\\
    \texttt{changchun.zou@ucf.edu}\\
  \end{minipage}
\end{center}
\vspace{1.0em}

\begin{center}
    {\large\sffamily\bfseries Abstract}
\end{center}
\vspace{0.15em}
\begin{center}
    \begin{minipage}{0.85\textwidth}
        Modern vehicles remain vulnerable to unauthorized use and theft despite traditional security measures including immobilizers and keyless entry systems. Criminals exploit vulnerabilities in Controller Area Network (CAN) bus systems to bypass authentication mechanisms, while social media trends have expanded auto theft to include recreational joyriding by underage drivers. Driver authentication via CAN bus data offers a promising additional layer of defense-in-depth protection, but existing open-access driver fingerprinting datasets suffer from critical limitations including reliance on decoded diagnostic data rather than raw CAN traffic, artificial fixed-route experimental designs, insufficient sampling rates, and lack of demographic information.\\
        
        This paper provides a comprehensive review of existing open-access driver fingerprinting datasets, analyzing their strengths and limitations to guide practitioners in dataset selection. We introduce the Kidmose CANid Dataset (KCID), which addresses these fundamental shortcomings by providing raw CAN bus data from 16 drivers across four vehicles, including essential demographic information and both daily driving and controlled fixed-route data. Beyond dataset contributions, we present a driver authentication anti-theft framework and implement a proof-of-concept prototype on a single-board computer. Through live road trials with an unaltered passenger vehicle, we demonstrate the practical feasibility of CAN bus-based driver authentication anti-theft systems. Finally, we explore diverse applications of KCID beyond driver authentication, including driver profiling for insurance and safety assessments, mechanical anomaly detection, young driver monitoring, and impaired driving detection. This work provides researchers with both the data and methodological foundation necessary to develop robust, deployable driver authentication systems and advance automotive behavioral biometrics research.
    \end{minipage}
\end{center}
\vspace{0.5em}

\textit{Keywords:} Driver fingerprinting, machine learning, automotive, cyber-physical systems, authentication, access control, anti-theft

\section{Introduction} \label{introduction}

Automobile theft (also known as ``auto theft,'' ``motor vehicle theft,'' or ``grand theft auto'') has evolved from mechanical lockpicking to sophisticated cyberattacks targeting a vehicle's electronic networks and systems. Contemporary car thieves circumvent anti-theft protections by exploiting the Controller Area Network (CAN) bus, gaining access through vulnerable external components such as headlight assemblies. Thieves inject fraudulent commands that mimic legitimate key authentication sequences, thereby defeating immobilizer protections and enabling unauthorized vehicle ignition and theft \cite{smart-vehicles-in-smart-cities,theft}.

Auto theft has proliferated and expanded beyond traditional criminal activity to include social media-driven challenges, where viral content demonstrates how to steal specific vehicle models. This phenomenon has resulted in alarming trends involving children and teenagers attempting auto theft for recreational purposes, frequently leading to traffic accidents and fatalities \cite{smart-vehicles-in-smart-cities,theft-tiktok}.

\subsection{Auto Theft} \label{auto-theft}

    This section examines contemporary auto theft. Rather than addressing traditional mechanical techniques, we focus on modern electronic exploitation strategies that target keyless entry systems, unprotected ignition systems, and the CAN bus. These are not necessarily new issues (unprotected ignition systems were prevalent before immobilizer technology); however, our interconnected digital ecosystem has enabled auto theft techniques to propagate through social media, resulting in an epidemic of theft and joyriding.
    
    \subsubsection{Keyless Entry}
    
        Research has identified critical vulnerabilities that facilitate unauthorized vehicle access and theft. Chen et al. \cite{iot-taxonomy-challenges-practice} demonstrated that criminals can exploit Passive Keyless Entry and Start (PKES) systems through relay attacks \cite{smart-vehicles-in-smart-cities}. These attacks intercept and relay communication signals between a vehicle and its key fob, enabling unauthorized access regardless of the encryption or authentication protocols implemented.
        
        Francillon, Danev, and Capkun \cite{relay-attack-passive-keyless-entry} conducted experimental relay attacks on ten vehicles from eight manufacturers, facilitating auto theft. Their research revealed that attackers could unlock and start vehicles even when the key fob was positioned up to 50 meters away. The attack relies on relay devices positioned strategically near both the target vehicle and the owner's key fob, with the required equipment costing between \$100 and \$1,000 \cite{smart-vehicles-in-smart-cities}.
        
        Yang et al. \cite{resisting-relay-attacks} identified the fundamental vulnerability: vehicles authenticate communication capability rather than physical proximity \cite{smart-vehicles-in-smart-cities}. This design assumption enables criminals to exploit relay devices that extend the communication range beyond the intended security perimeter.
    
    \subsubsection{Ignition}
    
        Certain Kia and Hyundai models are particularly vulnerable to vehicle theft. While 96\% of vehicles manufactured between 2015 and 2019 include electronic immobilizers, many late-model Kia and Hyundai vehicles lack this protection \cite{theft-no-insurance,smart-vehicles-in-smart-cities}. Thieves can break a rear window, remove steering column panels, and use simple tools such as USB cables to start the ignition and drive away \cite{theft-tiktok,smart-vehicles-in-smart-cities}.
        
        This vulnerability spread through social media in 2020, leading to an auto theft epidemic. In Milwaukee, over half the car thieves caught were children \cite{theft-milwaukee,smart-vehicles-in-smart-cities}. The epidemic resulted in at least fourteen crashes and eight fatalities according to the U.S. National Highway Traffic Safety Administration \cite{theft-nhtsa,smart-vehicles-in-smart-cities}. These tragic deaths occurred in large part because inexperienced children and teens were able to gain access to motor vehicles without supervision. With appropriate anti-theft measures, such as immobilizers and driver authentication, many of these deaths can be prevented.
        
        Even immobilizer-equipped vehicles remain vulnerable to sophisticated attacks. Most immobilizers employ Radio Frequency Identification (RFID) transponders embedded within physical keys. These transponders or ``chips'' store authentication secrets. During ignition, the vehicle challenges the key's transponder; if the transponder does not reply with the secret, the engine will not start. However, Garcia et al. \cite{lock-it-and-still-lose-it} observed that most immobilizers have historically relied on weak, proprietary cryptography---perhaps due to cost constraints and the computational limitations of RFID chips. Several cryptanalytic breakthroughs have compromised various widely-used immobilizers: the DST40 cipher in Texas Instruments' Digital Signature Transponder (2005) \cite{dst-immobilizer-broken}, NXP's Hitag2 transponders (2012, with bypass times under six minutes) \cite{hitag2-immobilizer-broken}, and the Megamos Crypto transponder (2013) \cite{megamos-immobilizer-broken}. As a result, numerous RFID immobilizers can be cloned, effectively negating their anti-theft protection \cite{lock-it-and-still-lose-it}.
    
    \subsubsection{The CAN Bus}
    
        As manufacturers implement countermeasures, criminals hone newer, more sophisticated auto theft techniques. Tindell \cite{theft} identified attacks where thieves accessed the CAN bus by attacking exterior components---e.g., headlights or paneling---to reach in-vehicle network cables. Once connected, attackers sent messages mimicking legitimate key authentication, telling the vehicle to unlock the doors and start the engine, allowing them to drive off with the vehicle. This vulnerability exists because the CAN bus, developed in 1983, lacks authentication, authorization, and encryption \cite{smart-vehicles-in-smart-cities}.

\subsection{Research Gap and Contribution}

    While several papers have reviewed open-access CAN intrusion detection datasets, no comprehensive review exists for driver fingerprinting datasets, which facilitate driver authentication. This paper addresses that gap by (1) providing practitioners with guidance on existing driver fingerprinting datasets and (2) introducing a novel dataset that overcomes critical limitations in existing resources.
    
    \begin{itemize}
        \item \textbf{Problem Statement:} Modern vehicles remain vulnerable to unauthorized use and auto theft despite existing security measures. Often, driver authentication research relies on datasets with fundamental methodological limitations that prevent the development of robust, deployable systems.
        
        \item \textbf{Motivation:} Robust driver authentication mechanisms can prevent both ``grand theft auto'' by criminal enterprises and unauthorized use by teenage or underage drivers. However, the driver fingerprinting datasets currently available have significant limitations, especially their dependence on unrealistic driver authentication scenarios. These limitations hinder researchers' efforts to develop generalizable driver authentication systems.
        
        \item \textbf{Research Gap:} Existing datasets suffer from critical limitations including reliance on decoded diagnostic data rather than raw CAN bus traffic, artificial fixed-route experimental designs, insufficient sampling rates, inadequate data volume, limited vehicle diversity, and lack of information about demographic diversity. These limitations severely constrain the development of driver authentication systems capable of (1) distinguishing between drivers with similar demographic profiles and (2) generalizing across different vehicle types and user populations.
        
        \item \textbf{Contribution:} This paper provides a comprehensive overview of existing driver authentication datasets and their limitations. We introduce the \textit{Kidmose CANid Dataset (KCID),} which addresses these fundamental shortcomings by (1) providing raw CAN bus data collected from multiple drivers across several different vehicles and by (2) including essential demographic information. Our dataset enables researchers to develop and evaluate driver authentication systems under realistic conditions, assess performance across diverse demographic groups, and create generalizable solutions suitable for real-world deployment. Finally, we develop a driver authentication anti-theft framework and implement a proof-of-concept prototype on a single-board computer. Through live road trials with an unaltered passenger vehicle, we demonstrate the practicability of our prototype and our overall approach.
    \end{itemize}

\subsection{Organization}

    The remainder of this paper is organized as follows: Section \ref{background-and-related-work} reviews existing literature on driver fingerprinting and authentication, enumerates existing open-access driver fingerprinting datasets, and outlines the differences between raw CAN bus data and decoded OBD-II data. Section \ref{kcid} describes our new dataset---\textit{The Kidmose CANid Dataset (KCID)}---including the data collection process, the vehicles, the drivers, and the routes and driving conditions. Section \ref{applications} discusses practical applications of driver fingerprinting datasets---from driver authentication to fleet management. Section \ref{anti-theft-framework-and-proof-of-concept} describes our driver authentication anti-theft framework and proof of concept. Section \ref{limitations-future-work} identifies limitations of our work and highlights future research directions. Section \ref{conclusion} concludes our work.

\section{Background \& Related Work} \label{background-and-related-work}

This section examines existing driver fingerprinting and authentication methodologies, analyzes open-access driver fingerprinting datasets available to researchers, and outlines the advantages and drawbacks of raw CAN bus data versus decoded diagnostic data for driver authentication applications.

\subsection{Driver Fingerprinting and Authentication} \label{driver-fingerprinting-and-authentication}
    
    Authenticating legitimate vehicle owners and users began with physical car keys. Simple mechanical keys unlock doors and operate ignition locks to start engines. These keys are vulnerable to duplication through temporary access to the originals or even high-quality photographs. Consequently, immobilizers were integrated with mechanical keys, creating ``chip keys.'' Most immobilizers employ Radio Frequency Identification (RFID) transponders embedded within physical keys. These transponders contain authentication secrets. During ignition, the vehicle challenges the transponder for the secret; without the correct response, the engine will not start. This technology significantly reduces theft \cite{lock-it-and-still-lose-it}. However, some modern vehicles produced between 2015 and 2019 lack immobilizers, leaving them substantially more vulnerable \cite{smart-vehicles-in-smart-cities,theft-no-insurance}.
    
    As discussed in Section \ref{auto-theft}, immobilizers can be circumvented---both by underage joyriders and professional criminals. Keyless entry systems and in-vehicle networks, especially Controller Area Networks, are also vulnerable \cite{smart-vehicles-in-smart-cities}. Therefore, a defense-in-depth approach is warranted. Researchers have turned to driver fingerprinting and authentication as an additional layer of anti-theft protection. Driver fingerprinting identifies drivers based on their driving data, and driver authentication differentiates between legitimate drivers and unauthorized drivers (e.g., would-be car thieves).
    
    In 2016, Enev et al. \cite{driver-fingerprinting} demonstrated that driver fingerprinting poses a significant threat to privacy, as drivers can be readily identified from diagnostic data. They collected diagnostic data via OBD-II queries (see Section \ref{can-bus-data-vs-diagnostic-data} for the differences between diagnostic data and raw CAN data). Ultimately, they recorded 16 different diagnostic parameters---including brake pedal position, steering wheel angle, and lateral acceleration---for 15 different drivers. All participants drove a 2009 sedan, performing maneuvers in an isolated parking lot before navigating a defined 50-mile loop through the Seattle metropolitan area. Enev et al. implemented four machine learning algorithms---support vector machine, random forest, naïve Bayes, and $k$-nearest neighbors---to differentiate drivers in a binary, pairwise fashion. Using all available sensors and 90\% of the collected diagnostic driving data, the authors achieved 100\% accuracy in differentiating all 15 drivers.
    
    Later in 2016, Kwak, Woo, and Kim \cite{hcrl-driving-dataset} collected diagnostic data from 10 drivers operating a Kia Soul on a fixed route in Seoul, South Korea, creating the HCRL Driving Dataset \cite{hcrl-driving-dataset-DATA} (see Section \ref{hcrl-driving-dataset}). They subdivided the driving data by road type and driving conditions: (1) parking lot, (2) city street (with traffic signals and pedestrian crossings), and (3) controlled-access expressway. Following several feature engineering steps, they employed four machine learning algorithms for driver identification: decision tree, random forest, $k$-nearest neighbors, and multi-layer perceptron. Unlike Enev et al. \cite{driver-fingerprinting}, they trained exclusively on the legitimate drivers' data. All four algorithms achieved accuracies exceeding 90\% across all three road types. Accuracy was generally higher on city streets and expressways than parking lots. The random forest model achieved the highest accuracy values: 99.8\%, 99.8\%, and 99.3\% for city streets, expressways, and parking lots, respectively.
    
    Similarly, in 2019, Kang, Park, and Kim \cite{auto-theft-detection-clustering} proposed a driver authentication scheme that depended only on ``owner'' data. Their approach employed $k$-means clustering to cluster key features from the owner's driving data, then used reconstruction error with a suitable threshold to distinguish the owner from potential thieves during new driving sessions. Like Enev et al. \cite{driver-fingerprinting}, they collected diagnostic data via OBD-II queries rather than raw CAN data. Like Kwak, Woo, and Kim \cite{hcrl-driving-dataset}, they trained without unauthorized driver (thief) data. Unlike Kwak, Woo, and Kim \cite{hcrl-driving-dataset}, Kang, Park, and Kim's scheme \cite{auto-theft-detection-clustering} accommodates only one authorized driver (owner) per vehicle. Their experimental results yielded model accuracies of at least 97\% to 99\% for key features, demonstrating the scheme's viability as a theft detection solution.
    
    Also in 2019, Park and Kim \cite{this-car-is-mine-dataset,this-car-is-mine-dataset-escar} collected diagnostic driving data from four drivers operating a Hyundai YF Sonata, creating the This Car is Mine! Dataset (see Section \ref{this-car-is-mine-dataset}). All drivers followed South Korean traffic laws. Following feature engineering and sliding window implementation to enhance training data, they developed a driver authentication scheme based on generative adversarial networks (GANs), specifically recurrent GANs (RGANs). In RGANs, both discriminator and generator are recurrent neural networks (RNNs). Like several previous approaches, they trained exclusively on ``owner'' data, with the discriminator distinguishing between the owner's actual driving data and imitation data produced by the generator. They conducted four experiments with different drivers designated as the ``owner,'' treating all others as thieves. Park and Kim's GAN-based driver identification model attained an average accuracy of 88.4\%, an average F1-score of 78.9\%, and successfully learned owner driving patterns from just 33 minutes of driving data.
    
    In 2021, Ahmadi-Assalemi et al. \cite{uk-driving-dataset} developed the UK Driving Dataset: 19 Drivers, featuring 19 drivers operating the same 2009 Mercedes Benz CLS passenger vehicle on an identical route in the London metropolitan area. They collected demographic information including age, gender, and ethnicity, then they collected diagnostic data via OBD-II queries during driving sessions. During feature selection, they prioritized features universally available in OBD-II data, such as torque and revolutions per minute (RPMs). Using a random forest classifier implemented in R, they designated three drivers as authorized (representing a family) and 16 as unauthorized. With only 10 seconds of driving data, they achieved 99.7\% accuracy at 95\% confidence. They observed notable differences in driving patterns between male and female drivers, enabling sex-based differentiation. Male drivers exhibited significantly greater fluctuation in longitudinal acceleration, RPMs, vehicle speed, and torque both during and between laps compared to female drivers. Classifier performance also varied by driver sex.
    
    In 2023, Khan, Lim, and Kim \cite{theft-detection-can-bus} developed a driver classification intrusion detection system (IDS) using an LSTM-FCN model, which combines the strengths of fully convolutional networks (FCNs) and long short-term memory (LSTM). They evaluated the model using the HCRL Driving Dataset \cite{hcrl-driving-dataset,hcrl-driving-dataset-DATA} and the This Car is Mine! Dataset \cite{this-car-is-mine-dataset,this-car-is-mine-dataset-escar,this-car-is-mine-dataset-DATA}, achieving accuracies of 99.36\% and 96.37\%, respectively. Like all previous approaches, they utilized diagnostic data rather than raw CAN data.
    
    For additional driver fingerprinting and authentication schemes based on decoded diagnostic data, we refer the reader to \cite{whos-driving-my-car,driving-behavior-uncontrolled,barreto-obd-ii-datasets,this-car-is-mine-dataset,this-car-is-mine-dataset-escar,human-behavior-driving,live-di,lightweight-driver-behavior,driver-identification-only-can,vehicle-theft-gan}.
    
    Lestyán et al. \cite{extracting-signals} adopted a different approach to driver authentication by collecting raw CAN data rather than diagnostic data. They gathered raw CAN bus data from eight vehicles driven by 33 drivers, employing various machine learning techniques to extract signals including brake pedal position, accelerator pedal position, clutch position, engine RPMs, and speed. They developed random forest classifiers that, when trained on a base vehicle with known target signals, could reliably identify three signals (accelerator pedal position, engine RPM, and speed) in other vehicles. They excluded brake pedal position and clutch position from random forest-based extraction, as these signals were not present across all vehicles. Using four extracted signals---brake pedal position, accelerator pedal position, engine RPM, and speed----they attempted driver re-identification for the 2018 Opel Astra. Through 10-fold cross-validation, they randomly selected five drivers and constructed binary classifiers for each driver pair, achieving 77\% average precision and demonstrating the approach's potential.

\subsection{Open-Access Driver Fingerprinting Datasets} \label{open-access-driver-fingerprinting-datasets}
    
    \begin{table*}[htb!]
    \footnotesmallb
    \centering
        \begin{threeparttable}
        \caption{Comparative Summary of Existing Open-Access Driver Fingerprinting Datasets and KCID} \label{tab:dataset-comparison}
        
        \begin{tabular}{|>{\raggedright\arraybackslash}p{3.0cm}|>{\raggedright\arraybackslash}p{0.8cm}|>{\raggedright\arraybackslash}p{3.4cm}|>{\raggedright\arraybackslash}p{1.1cm}|>{\raggedright\arraybackslash}p{1.7cm}|>{\raggedright\arraybackslash}p{1.7cm}|>{\raggedright\arraybackslash}p{1.5cm}|} \hline
        \rowcolor{LightSteelBlue3!80}
            \textbf{Dataset} &
            \textbf{Year\tnote{1}} & \textbf{Vehicle(s)} & \textbf{Drivers} &
            \textbf{Data Type} & \textbf{Route Type} & \textbf{Frequency} \\ \hline \hline
            
            \rowcolor{LightSteelBlue3!40}
            HCRL Driving Dataset \cite{hcrl-driving-dataset,hcrl-driving-dataset-DATA} &
            2016 & 1 vehicle: \newline KIA Soul\tnote{2} & 10 &
            Decoded OBD-II data & Fixed route & 1 Hz \\ \hline
            
            \rowcolor{LightSteelBlue3!10}
            Barreto dailyRoutes \cite{barreto-obd-ii-datasets,barreto-obd-ii-datasets-DATA-1} &
            2018 & 14 different vehicles: \newline 2003 - 2016 model years & 14 &
            Decoded OBD-II data & Daily driving & 0.14 Hz \\ \hline
            
            \rowcolor{LightSteelBlue3!40}
            Barreto 19drivers \cite{barreto-obd-ii-datasets,barreto-obd-ii-datasets-DATA-2} &
            2018 & 1 vehicle: \newline 2015 Chevrolet S10 & 19 &
            Decoded OBD-II data & Fixed route & 0.14 Hz \\ \hline
            
            \rowcolor{LightSteelBlue3!10}
            AEGIS Big Data Project \cite{aegis} &
            2019 & 1 vehicle: \newline unknown model & 3 &
            Decoded OBD-II data + GPS data & Not specified & 20 Hz \\ \hline
            
            \rowcolor{LightSteelBlue3!40}
            This Car is Mine! \cite{this-car-is-mine-dataset,this-car-is-mine-dataset-escar,this-car-is-mine-dataset-DATA} &
            2019 & 1 vehicle: \newline Hyundai YF Sonata & 4 &
            Decoded OBD-II data & Fixed route & 1 Hz \\ \hline

            \rowcolor{LightSteelBlue3!10}
            UK Driving Dataset: \newline 19 Drivers \cite{uk-driving-dataset} &
            2021 & 1 vehicle: \newline 2009 Mercedes Benz CLS & 19 &
            Decoded OBD-II data & Fixed route & 1 Hz \\ \hline

            \rowcolor{LightSteelBlue3!40}
            Kidmose CANid Dataset (KCID)\tnote{3} &
            2025 & 4 different vehicles: \newline 2011 Chevrolet Traverse, \newline 2017 Ford Focus, \newline 2017 Subaru Forester, \newline 2022 Honda CR-V Touring & 16 &
            Raw CAN data & Daily driving, fixed route & Varies\tnote{4} \\ \hline
        \end{tabular}
        \footnotesize
        \begin{tablenotes}
            \item [1] Publication year of the peer-reviewed paper introducing the dataset, if applicable, or year of initial dataset publication.
            \item [2] According to the dataset metadata.
            \item [3] See Section \ref{kcid}.
            \item [4] Varies depending on the vehicle and driving conditions. Ranges from ~1,000 Hz to ~2,500 Hz for the Traverse, Focus, and Forester.
        \end{tablenotes}
        \end{threeparttable}
    \end{table*}

    Driver authentication via CAN bus data has emerged as a promising approach for enhancing vehicle security and preventing unauthorized access. Several open-access datasets have been published to facilitate research in this domain, each with distinct characteristics, strengths, and limitations. This section provides a comprehensive analysis of existing datasets to contextualize the research gaps that our proposed Kidmose CANid Dataset (KCID) addresses.

    Table \ref{tab:dataset-comparison} presents an analysis and summary of existing open-access driver fingerprinting datasets available to automotive security researchers and practitioners, alongside our novel Kidmose CANid Dataset (KCID) for comparison. We focus specifically on datasets suitable for driver identification, fingerprinting, profiling, and authentication. Minimum criteria include two or more labeled drivers, ideally with multiple drivers operating the same vehicle. We exclude datasets that are not publicly available as of this writing. We successfully downloaded and analyzed all six existing open-access driver fingerprinting datasets shown in Table \ref{tab:dataset-comparison}.

    \subsubsection{HCRL Driving Dataset} \label{hcrl-driving-dataset}
    
        To our knowledge, the HCRL Driving Dataset \cite{hcrl-driving-dataset,hcrl-driving-dataset-DATA} constitutes the earliest open-access driver fingerprinting dataset. Kwak, Woo, and Kim collected data from 10 drivers who each completed two roundtrips using the same vehicle---a KIA Soul (2016 or earlier model). All participants navigated a fixed route connecting Korea University and SANGAM World Cup Stadium. To control for traffic conditions, all experiments were conducted between 8:00 p.m. and 11:00 p.m. on weekdays.
        
        The approximately 29-mile (46-kilometer) roundtrip route included three distinct driving environments: (1) city streets with traffic signals and pedestrian crossings, (2) controlled-access expressways, and (3) parking lots. The authors reported collecting approximately 23 hours of driving data, averaging 2.3 hours per participant. Our analysis revealed 94,380 data entries in the dataset. Since data was collected per second, this represents over 26 hours of driving data, or approximately 2.6 hours per participant---sufficient for most machine learning applications.
        
        Due to incompatibilities between South Korea's national security laws and Google Maps requirements for advanced map services, Google Maps does not currently provide driving directions in South Korea. Other online map providers do not allow users to create and share custom driving routes. We refer practitioners to the original paper \cite{hcrl-driving-dataset}, where the driving route is overlaid on satellite imagery.
        
        The dataset provides decoded OBD-II data including parameters such as \texttt{Fuel\_consumption}, \texttt{Accelerator\_Pedal\_value}, and \texttt{Throttle\_position\_signal} with their corresponding numerical values. The dataset does not provide raw CAN data.
        
        While this dataset provides a foundation for driver fingerprinting research, several limitations constrain its applicability:
        
        \begin{itemize}
            \item Data was collected via diagnostic queries at a rate of 1 Hz, resulting in significant information loss compared to raw CAN bus traffic.
            \item The diagnostic query approach introduces latency as requests must be sent and responses received, potentially missing transient driving behaviors.
            \item The fixed route design creates an unrealistic scenario for distinguishing authorized from unauthorized drivers, as attackers would rarely follow exactly the same path as the legitimate owner.
        \end{itemize}

        Numerous driver identification and authentication schemes have leveraged the HCRL Driving Dataset for development and evaluation, including \cite{hcrl-driving-dataset,whos-driving-my-car,human-behavior-driving,live-di,lightweight-driver-behavior,driver-identification-only-can,vehicle-theft-gan,theft-detection-can-bus}.
        
        Link to the HCRL Driving Dataset: \href{https://ocslab.hksecurity.net/Datasets/driving-dataset}{\nolinkurl{ocslab.hksecurity.net/Datasets/driving-dataset}}
    
    \subsubsection{Barreto OBD-II Datasets} \label{barreto-datasets}
    
        Barreto published two complementary datasets with different experimental setups, dubbed ``dailyRoutes'' and ``19drivers'' \cite{barreto-obd-ii-datasets,barreto-obd-ii-datasets-DATA-1,barreto-obd-ii-datasets-DATA-2}. Both datasets provide decoded diagnostic data collected via an ELM 327 Bluetooth device connected to the vehicle's OBD-II port.
        
        \textbf{Barreto dailyRoutes Dataset.} The dailyRoutes Dataset features fourteen different drivers operating fourteen different vehicles manufactured between 2003 and 2016. Data was collected by plugging an ELM 327 device into the OBD-II port and then connecting an Android device to the ELM 327 device via Bluetooth. This setup captured the drivers' normal daily driving routines. Major limitations include:
        
        \begin{itemize}
            \item Since each driver used a different vehicle, the dataset cannot be used to distinguish different drivers operating the same vehicle.
            \item The Bluetooth-based collection method introduces significant latency and packet loss \cite{id-sequence-android}.
            \item The approximately seven-second interval (0.14 Hz) between command sequences results in substantial loss of fine-grained driving behavior data.
        \end{itemize}
        
        Links to the Barreto dailyRoutes Dataset: \href{https://github.com/cephasax/OBDdatasets}{\nolinkurl{github.com/cephasax/OBDdatasets}} and \href{https://www.kaggle.com/datasets/cephasax/obdii-ds3}{\nolinkurl{www.kaggle.com/datasets/cephasax/obdii-ds3}}
        
        \textbf{Barreto 19drivers Dataset.} The 19drivers Dataset addresses several limitations of the dailyRoutes dataset by having nineteen different drivers operate the same vehicle---a 2015 Chevrolet S10---along the same route. Due to scheduling challenges, Barreto did not control for time of day, traffic, etc. One experienced driver was instructed to drive in four different styles: ``prudente'' (cautious), ``moderado'' (moderate), ``apressado'' (hurried), and ``imprudente'' (reckless).
        
        The route forms an 11.7-mile (18.8-kilometer) loop with a slightly different return path. The route is available on Google Maps: \href{https://maps.app.goo.gl/9uGkTbaMyb3aNgNN6}{\nolinkurl{maps.app.goo.gl/9uGkTbaMyb3aNgNN6}}
        
        Data was collected every seven seconds, yielding 8,261 data entries. This corresponds to 57,827 seconds (approximately 16 hours) of data, averaging about 51 minutes per driver. This constitutes significantly less data than the HCRL Driving Dataset, and the reduced granularity may limit applicability for data-intensive machine learning applications.
        
        The 19drivers Dataset offers several advantages:
        
        \begin{itemize}
            \item Eliminates vehicle-specific variables that could confound driver identification.
            \item Enables direct comparison between drivers under controlled conditions.
            \item Provides insight into how driving style affects identification accuracy.
        \end{itemize}
        
        However, a significant limitation remains; namely, the fixed route creates an artificial scenario inconsistent with real-world driver fingerprinting and authentication contexts. The fixed route could prove valuable for feature engineering and the early stages of model training. However, for the late stages of model training---as well as model testing---different routes are needed. When tackling the driver authentication problem, we cannot assume that an unauthorized driver (e.g., a thief) would follow the same route as an authorized driver (e.g., the vehicle owner). Additionally, as with the dailyRoutes Dataset, the 19drivers Dataset suffers from Bluetooth-induced latency and packet loss \cite{id-sequence-android}, as well as substantial information loss due to its 0.14 Hz sampling frequency.
        
        Link to the Barreto 19drivers Dataset: \href{https://github.com/cephasax/OBDdatasets}{\nolinkurl{github.com/cephasax/OBDdatasets}}
    
    \subsubsection{AEGIS Dataset}
    
        The AEGIS Dataset \cite{aegis,aegis-final-report,aegis-eu} is an example dataset from the AEGIS Big Data Project, a European Union-funded initiative exploring big data applications in the automotive domain. The dataset contains driving data from three drivers operating the same vehicle. The make and model of the vehicle are not specified. Similar to the HCRL Driving Dataset and the Barreto OBD-II Datasets, the AEGIS Dataset provides decoded diagnostic data rather than raw CAN traffic.
        
        One notable advantage of this dataset is its relatively high sampling frequency of 20 Hz, which preserves more fine-grained driving behavior information compared to other datasets in this domain. Additionally, OBD-II data is readily interpretable, thanks to its standardized nature.
        
        However, several significant limitations constrain the dataset's utility for driver authentication and fingerprinting research:
        
        \begin{itemize}
            \item The dataset provides only decoded OBD-II data, omitting the richer information available in raw CAN traffic.
            \item The single-vehicle approach limits generalizability, as researchers cannot assess whether developed methods will transfer to other vehicle makes and models.
            \item Documentation is extremely limited. The dataset is described merely as an ``example dataset'' representing a small fraction (922.1 MB) of the project's total data volume (104 GB across multiple automotive big data applications).
            \item Critical experimental details are missing, including driving routes, environmental conditions, traffic scenarios, and the distinction between urban versus highway driving contexts.
            \item The unspecified vehicle make and model complicates reproducibility and cross-study comparisons.
        \end{itemize}
        
        While the AEGIS Dataset contributes to the available resources for driver fingerprinting research, its limited documentation and scope significantly restrict its practical application for developing robust, generalizable authentication systems.
        
        Link to the AEGIS Dataset: \href{http://doi.org/10.5281/zenodo.3267184}{\nolinkurl{doi.org/10.5281/zenodo.3267184}}
    
    \subsubsection{This Car is Mine! Dataset} \label{this-car-is-mine-dataset}
        
        Published in 2019 and uploaded to IEEE Dataport in 2020, the This Car is Mine! Dataset features four drivers operating a single Hyundai YF Sonata \cite{this-car-is-mine-dataset,this-car-is-mine-dataset-escar,this-car-is-mine-dataset-DATA}. The experimental design establishes one participant as the legitimate ``owner'' while treating the remaining three as potential ``thieves,'' enabling evaluation of the authors' proposed driver authentication scheme. The authors' GAN-based driver identification model successfully learned the owner's driving pattern using only 33 minutes of driving data (see Section \ref{driver-fingerprinting-and-authentication}).
        
        The dataset contains decoded OBD-II data collected at 1 Hz sampling frequency, matching the HCRL Driving Dataset's approach. It includes 30 driving sessions distributed unevenly across participants (8, 8, 5, and 9 trips, respectively), with a limited feature set of 51 parameters representing a small subset of the data typically available in modern vehicles. All drivers followed an identical predetermined route, controlling for differences in driving conditions but reducing real-world applicability.
        
        Since data was collected per second, we examined the number of data entries per trip to estimate the average trip duration. Trips ranged from 1,716 to 2,296 seconds (with one outlier at 1,444 seconds), meaning that trips averaged approximately 30 minutes. This yields approximately 2.5 hours of data from the driver with the fewest trips and approximately 4.5 hours from the driver with the most trips---sufficient for data-intensive machine learning applications. Since Google Maps does not provide driving directions in South Korea, we refer the practitioners to the original papers \cite{this-car-is-mine-dataset,this-car-is-mine-dataset-escar}, where the driving route is overlaid on a digital map screenshot.
        
        While the data volume is sufficient, several limitations reduce the dataset's utility:
        
        \begin{itemize}
            \item The 1 Hz sampling rate introduces significant information loss compared to the high-frequency nature of actual CAN bus communications.
            \item The constrained feature set limits the potential for comprehensive behavioral modeling.
            \item The fixed route design creates an artificial authentication scenario, as unauthorized drivers would rarely follow the same paths as legitimate vehicle owners.
        \end{itemize}
        
        Despite these limitations, the dataset's ``owner-vs-thieves'' experimental framework provides valuable insights into driver authentication performance under controlled conditions and serves as a useful benchmark for comparing different driver identification approaches.
        
        Several driver identification and authentication schemes have leveraged the This Car is Mine! Dataset for development and evaluation, including \cite{this-car-is-mine-dataset,this-car-is-mine-dataset-escar,theft-detection-can-bus}.
        
        Link to the This Car is Mine! Dataset: \href{https://dx.doi.org/10.21227/qar8-sd42}{\nolinkurl{dx.doi.org/10.21227/qar8-sd42}}
        
    \subsubsection{UK Driving Dataset: 19 Drivers} \label{uk-driving-dataset}

        To our knowledge, the UK Driving Dataset: 19 Drivers \cite{uk-driving-dataset,uk-driving-dataset-DATA} represents the most recent driver fingerprinting dataset by publication year. Uploaded to ResearchGate in 2018 and published in 2021, the dataset features volunteers recruited via email and social media. To be eligible, would-be volunteers had to meet the following criteria: (1) be over 18 years of age, (2) be in possession of a valid driver's license, and (3) be covered by comprehensive car insurance. Volunteer drivers had opportunities to familiarize themselves with the experiment vehicle before data collection, hopefully reducing the drivers' uncertainty and discomfort with an unfamiliar vehicle---as well as the corresponding bias and noise.

        Nineteen drivers operated the same 2009 Mercedes Benz CLS passenger vehicle along the same route in the London metropolitan area. The route consisted of a 2.7-mile loop, primarily urban with traffic lights, roundabouts, low-speed local roads, and medium-speed main roads (40 miles per hour). The route is available on Google Maps: \href{https://maps.app.goo.gl/4FBCsTzfcefHqK8KA}{\nolinkurl{maps.app.goo.gl/4FBCsTzfcefHqK8KA}}
        
        Each volunteer drove the route three times during a 45-minute session, generating 8.1 miles (13 kilometers) of driving data per participant. To ensure each driver faced roughly equivalent traffic conditions, data collection sessions were scheduled for times when traffic patterns were comparable. Notably, the authors also collected demographic information from participants, including age, gender, and ethnicity.

        The dataset provides decoded OBD-II data collected at 1 Hz sampling frequency using a BAFX A5--58GD-LWQN Bluetooth-enabled OBD-II scanning tool. The scanning tool was physically plugged into the vehicle's OBD-II port and then connected to an Android tablet via Bluetooth, similar to the Barreto OBD-II Datasets. During feature selection, the authors prioritized features universally available in automotive OBD-II data, such as torque and RPMs.

        During our analysis, we found a total of 36,223 data entries in the dataset. Since data was collected per second, this represents approximately 10 hours of driving data across all 19 drivers, averaging approximately 32 minutes per driver. This volume may prove insufficient for some data-intensive applications; datasets such as the HCRL Driving Dataset and the This Car is Mine! Dataset may be more suitable for such purposes.
        
        The following are additional limitations of the UK Driving Dataset: 19 Drivers:
        
        \begin{itemize}
            \item The 1 Hz sampling rate introduces significant information loss compared to the high-frequency nature of actual CAN bus communications.
            \item The Bluetooth-based collection method introduces significant latency and packet loss \cite{id-sequence-android}.
            \item The fixed route design creates an artificial authentication scenario, as unauthorized drivers would rarely follow the same paths as legitimate vehicle owners.
        \end{itemize}
        
        Several driver identification and authentication schemes have leveraged the UK Driving Dataset: 19 Drivers for development and evaluation, including \cite{uk-driving-dataset,optimizing-driver-id-adaboost}.
        
        Link to the UK Driving Dataset: 19 Drivers: \href{https://doi.org/10.13140/RG.2.2.14505.49765}{\nolinkurl{doi.org/10.13140/RG.2.2.14505.49765}}
    
    \subsubsection{Research Gaps and Dataset Limitations} \label{research-gaps-and-dataset-limitations}
    
        While existing driver authentication datasets have facilitated valuable research, they collectively exhibit several critical limitations that constrain the development of robust, deployable systems:
        
        \begin{enumerate}
            \item \textbf{Data Collection Methodology:} To date, all publicly available driver fingerprinting datasets rely on decoded OBD-II data rather than raw CAN messages, introducing significant latency and information loss. The diagnostic query-response approach fails to capture the continuous, high-frequency nature of actual in-vehicle communications. Lestyán et al. \cite{extracting-signals} recognized this limitation and collected raw CAN data from eight vehicles driven by 33 drivers, developing random forest classifiers for both signal extraction and driver re-identification. However, to our knowledge, their dataset is not publicly available, preventing other researchers from building upon their work without independent data collection efforts.
            
            \item \textbf{Insufficient Temporal Resolution:} Sampling rates as low as 0.14 Hz fundamentally fail to capture the fine-grained driving behaviors essential for reliable driver authentication. Automotive CAN buses operate at frequencies orders of magnitude higher, containing behavioral nuances lost in these coarse-grained datasets.
            
            \item \textbf{Unrealistic Experimental Design:} The prevalence of fixed-route experimental designs creates artificial scenarios inconsistent with practical authentication applications. Unauthorized drivers would rarely follow the same predetermined paths as legitimate vehicle owners; as such, route-dependent authentication features are unreliable for real-world deployment.
            
            \item \textbf{Limited Vehicle Diversity:} Most datasets focus on single vehicles, inhibiting the development of authentication systems capable of generalizing across different makes, models, production years, and vehicle types.
            
            \item \textbf{Absence of Demographic Information:} Only one dataset---the UK Driving Dataset: 19 Drivers---provides demographic information about participating drivers. Demographic factors such as age, gender, ethnicity, driving experience, etc. can significantly influence driving behaviors and patterns.
        \end{enumerate}

        When it comes to driver fingerprinting and authentication, the absence of demographic information can be particularly limiting. Insurance industry data clearly demonstrates that driver age and gender correlate strongly with driving behavior and risk patterns \cite{fatality-facts,age-gender-forbes}. Male drivers are statistically more likely to engage in risky driving behaviors, including speeding and driving while impaired \cite{fatality-facts,female-drivers-safer}. Teenage drivers, due to inexperience, exhibit greater tendencies toward speeding and maintaining shorter following distances, along with higher rates of distracted driving, with 39\% of U.S. high school-age drivers reporting texting while driving in the past 30 days \cite{teen-risk-cdc}. Fatal crash involvement rates quantify these differences, with male drivers exhibiting rates three times higher than female drivers (5.3 versus 1.7 per 10,000 drivers) and the highest rates occurring among teenage drivers \cite{fatality-facts}.
        
        Ahmadi-Assalemi et al. \cite{uk-driving-dataset} observed these demographic differences directly when they developed and analyzed the UK Driving Dataset: 19 Drivers. They identified notable differences in driving behavior and patterns between male and female drivers, with male drivers exhibiting significantly greater fluctuation in longitudinal acceleration, RPMs, vehicle speed, and torque both during and between laps compared to female drivers. Moreover, the performance of their proposed driver identification scheme varied depending on the sex of the driver, demonstrating that demographic factors can influence not only driving patterns but also the effectiveness of driver identification and authentication systems.

        These findings have serious implications, especially with respect to driver authentication system design, development, and evaluation. Distinguishing between drivers with markedly different profiles---such as a young male driver versus an older female driver---may prove relatively straightforward due to the expected behavioral differences. However, robust driver authentication systems must reliably differentiate between drivers with similar demographic profiles, such as two young male drivers or two middle-aged female drivers. Without demographic information in existing datasets, researchers cannot assess whether their driver authentication systems achieve this critical level of differentiation capability.
        
        Furthermore, the demographic composition of training data significantly impacts system performance across different user populations. Driver authentication systems trained predominantly on data from young female drivers may exhibit reduced accuracy when deployed for older male drivers, highlighting the importance of demographically diverse training datasets for developing generalizable systems.
        
        These fundamental limitations collectively highlight the urgent need for a more comprehensive dataset that addresses these methodological and representational shortcomings. Such a dataset would provide researchers with the necessary foundation to develop robust, real-world driver authentication systems capable of reliable deployment across diverse vehicle types, driving scenarios, and user demographics.

\subsection{CAN Bus Data vs. Diagnostic Data} \label{can-bus-data-vs-diagnostic-data}

\lstset{
    tabsize=2,
    basicstyle=\ttfamily,
    showstringspaces=false,
    breaklines=true,
    moredelim=**[is][\color{red}]{@}{@},
    moredelim=**[is][\color{ForestGreen}]{\$}{\$},
    moredelim=**[is][\color{blue}]{+}{+},
    moredelim=**[is][\color{NavyBlue}]{\%}{\%},
}

\begin{lstfloat}[b!]
\caption{In-vehicle network traffic captures. Vehicle: 2011 Chevrolet Traverse. Driver: \texttt{female-all-ages-5}.}
\begin{sublstfloat}[t]{0.48\textwidth}
\footnotesize
\begin{lstlisting}
(1744044851.673900) can0 348#07AC07AA
(1744044851.674150) can0 17D#04E000007D000100
(1744044851.674400) can0 17F#0000000000000000
(1744044851.674600) can0 34A#07A507A3
(1744044851.674800) can0 2D1#030000000000
(1744044851.677850) can0 0F1#34070040
(1744044851.678150) can0 1EB#018C
(1744044851.678450) can0 0C7#01CD519E
(1744044851.678700) can0 0F9#01694006AB531912
(1744044851.678900) can0 189#0FFF0FFF30011912
(1744044851.679150) can0 199#0FFF0E70F19000FF
(1744044851.680550) can0 1F3#0020
(1744044851.681050) can0 0C1#105F36E6106593A7
(1744044851.681300) can0 0C5#10244FBB1025E9AE
(1744044851.681500) can0 1E5#46FFBFCED4003F00
(1744044851.681750) can0 0C9#8020591624000000
(1744044851.682000) can0 191#075B075B075B2493
(1744044851.682250) can0 1ED#41740666074E0800
(1744044851.682400) can0 1EF#000009F6
(1744044851.682650) can0 1A1#0000414000002400
(1744044851.682900) can0 1C3#075B074900000000
(1744044851.683550) can0 2C3#08FA067706777A00
(1744044851.684700) can0 1CB#100D00
(1744044851.687450) can0 19D#00000000001DD32D
(1744044851.687650) can0 1AF#000020
(1744044851.687900) can0 1F5#4404000400000900
(1744044851.688150) can0 0F1#00070040
(1744044851.690550) can0 1EB#0159
(1744044851.690850) can0 0C7#01E05D97
(1744044851.691050) can0 0F9#01694006AA533212
(1744044851.691300) can0 0C1#2062402720689CE3
\end{lstlisting}
\caption{Raw CAN traffic.}
\label{lst:raw-can-traffic-capture}
\end{sublstfloat}
\hfill
\begin{sublstfloat}[t]{0.48\textwidth}
\footnotesize
\begin{lstlisting}
(1751066848.332750) can0 2D1#030000000000
(1751066848.332950) can0 3FD#006666
(1751066848.336500) can0 19D#C0003FFD001E2717
(1751066848.336650) can0 1AF#000020
(1751066848.336900) can0 1F5#4606000400000900
$(1751066848.336901) can0 7DF#02010C5555555555$
(1751066848.337900) can0 0C1#00D17A99021B5057
(1751066848.338100) can0 0F1#00070040
(1751066848.338300) can0 0C5#023AB8F200DEEE16
(1751066848.338550) can0 184#0002000001FE
(1751066848.338750) can0 1C7#0FFFAFFF03FF3F
(1751066848.338950) can0 1CD#07FF08017F
(1751066848.339150) can0 1E5#46FFF4AF9C000B00
(1751066848.339400) can0 1E9#0003000C00030000
(1751066848.339650) can0 0C9#801DF41304000000
(1751066848.339900) can0 191#076207AC075C040F
(1751066848.340050) can0 0C7#037EAE52
(1751066848.340300) can0 0F9#00BD400CDA356421
(1751066848.340400) can0 1EB#0157
(1751066848.340650) can0 189#CFFF0FFF2FFE6421
(1751066848.340900) can0 1ED#413F08BB07E80800
(1751066848.341150) can0 199#CFFF0E70F18D00FF
(1751066848.341300) can0 1EF#00000A0C
(1751066848.341500) can0 2C3#08E5065006509F00
%(1751066848.347550) can0 7E8#04410C1E47000000%
(1751066848.347800) can0 0C1#10D7841F12225B78
(1751066848.348000) can0 0C5#1240C27D10E5F93A
(1751066848.348200) can0 0F1#1C070040
(1751066848.348400) can0 1E5#46FFF2CF38000C00
(1751066848.349500) can0 0C9#801DC81603000000
(1751066848.349750) can0 191#075707A70751030C
\end{lstlisting}
\caption{CAN traffic containing an OBD-II query (\color{ForestGreen}green\color{black}) and an OBD-II response (\color{NavyBlue}blue\color{black}).}
\label{lst:obd-ii-query-response}
\end{sublstfloat}
\end{lstfloat}

    The choice between raw CAN bus data and decoded diagnostic data fundamentally impacts driver authentication research. Though similar, both options have advantages and trade-offs. Figures \ref{lst:raw-can-traffic-capture} and \ref{lst:obd-ii-query-response} highlight these differences using traffic captures from KCID's 2011 Chevrolet Traverse driven by \texttt{female-all-ages-5}. Figure \ref{lst:raw-can-traffic-capture} exclusively captures native CAN communications---the continuous stream of messages exchanged between ECUs during normal vehicle operation. Figure \ref{lst:obd-ii-query-response} demonstrates the diagnostic OBD-II query-response mechanism: an external device transmits a request (arbitration ID \texttt{0x7DF}, shown in \textcolor{ForestGreen}{green}) querying engine RPMs (PID \texttt{0x0C}), and the vehicle responds (arbitration ID \texttt{0x7E8}, shown in \textcolor{NavyBlue}{blue}) with the requested parameter value. Most existing driver authentication datasets (Section \ref{open-access-driver-fingerprinting-datasets}) provide only the decoded responses; the surrounding CAN traffic and the diagnostic requests themselves are filtered out, resulting in substantial information loss.

    \subsubsection{Advantages of Raw CAN Bus Data} \label{advantages-of-raw-can-bus-data}

        Raw CAN bus data offers several compelling advantages over diagnostic data collected via OBD-II queries and subsequently decoded:
        
        \textbf{Data Coverage.} Diagnostic data collection is constrained by standardized protocols and predetermined queries, capturing a limited subset of the vehicle's data. Raw CAN data provides access to all ECU communications that were transmitted across the CAN bus, revealing behavioral patterns invisible to diagnostic approaches.
        
        \textbf{Temporal Coverage.} Existing driver authentication datasets collect diagnostic data at 0.5-, 1-, or even 7-second intervals. In many modern vehicles, the CAN bus handles thousands of transmissions in just one second. Raw CAN bus data captures high-frequency information at native temporal resolution, preserving the transient behaviors and rapid control adjustments that characterize individual driving styles.
        
        \textbf{Fidelity.} Diagnostic data collection---especially when provided by specialized repair shop tools---often involves filtering, converting, and aggregating data, meaning that some fine-grained data is lost. Raw CAN bus data maintains the original precision and the original data volume, ensuring subtle behavioral differences are preserved rather than averaged out.

    \subsubsection{Challenges of Raw CAN Bus Data}
    
        It can be difficult to use raw CAN bus data to its full potential, for several reasons:

        \textbf{Signal Relevance.} Raw CAN traffic contains substantial noise---e.g., periodic status messages and ECU communications unrelated to driving behavior. This irrelevant information can overwhelm authentication-relevant signals. Researchers can employ several strategies to address this challenge. Feature engineering and filtering techniques can isolate behaviorally relevant messages. Machine learning approaches can automatically identify authentication-relevant features while suppressing noise, maximizing the utility of raw CAN data. For example, principal component analysis (PCA) facilitates dimensionality reduction, while clustering algorithms (e.g., $k$-means) and autoencoders enable feature learning. We refer the reader to \cite{pca-data-exploration,anomaly-detection-statistics,autoencoder-anomaly-detection,driver-fingerprinting,driver-identification-only-can,hcrl-driving-dataset,can-fp,all-learning-survey,classification-ids,lstm-4,deep-learning-time-series-survey,anomaly-detection-time-series,deep-learning-time-series} and Section \ref{machine-learning} for more information.
        
        \textbf{Interpretability.} Unlike decoded diagnostic data with standardized, human-readable parameters, raw CAN traffic consists of manufacturer-specific hexadecimal frames that require reverse engineering to extract meaningful behavioral features. However, recent research has developed efficient, automated and semi-automated CAN bus reverse engineering strategies that address this challenge. Signal extraction techniques can identify and isolate specific vehicle parameters (e.g., speed, throttle position) from undocumented CAN traffic, enabling feature-based authentication approaches. We refer the reader to \cite{extracting-signals,can-reverse-engineering,recan-dataset,classification-ids}.

\section{The Kidmose CANid Dataset (KCID)} \label{kcid}

The Kidmose CANid Dataset (KCID) contains CAN bus data collected by Brooke and Andreas \textbf{\textit{Kidmose}} from 16 drivers across four vehicles. Many of the participating drivers are recognized in the Acknowledgments (Section \ref{acknowledgments}). The term ``CANid'' reflects the dataset's source and purpose: data collected from the \textbf{\textit{CAN}} bus for driver \textbf{\textit{id}}entification research. This section describes our data collection hardware, configuration, and methodology, as well as KCID's composition and structure, ensuring data quality and research reproducibility.

\subsection{Data Collection}
    
    Initial experiments with low-cost hardware revealed fundamental limitations that compromise data quality and research validity, leading us to select equipment that meets our specific research requirements.
    
    \subsubsection{Rejected Hardware}
    
        We initially evaluated an ELM 327 Bluetooth device as a low-cost data collection option. This device \textit{physically} connects to a vehicle's OBD-II port and \textit{wirelessly} connects to a Bluetooth-enabled smartphone or tablet, which can display, log, and store the data. It costs approximately 20 USD at various online retailers, including Amazon.com. However, ELM 327 devices suffer from critical limitations that make them unsuitable for research applications \cite{id-sequence-android}, specifically:
        
        \begin{itemize}
            \item \textbf{Indirect CAN bus access.} The device does not tap directly into CAN bus traffic. Instead, it sends diagnostic queries and processes the vehicle's responses, introducing communication delays and data gaps.
            \item \textbf{Limited buffer capacity \& limited Bluetooth transmission speeds.} The device receives and buffers CAN data before sending it onward to a Bluetooth-enabled smartphone or tablet. If the device is receiving data faster than it can transmit data, then the buffer will fill up, and data will be lost.
            \item \textbf{Reduced data fidelity.} The combination of communication delays, data gaps, and buffering limitations compromise the data's temporal accuracy and overall fidelity.
        \end{itemize}
        
        These limitations affect existing datasets that rely on ELM 327 devices, including the Barreto OBD-II datasets and the UK Driving Dataset: 19 Drivers described in Sections \ref{barreto-datasets} and \ref{uk-driving-dataset}, respectively. Based on these findings, we sought more reliable equipment that could better meet our research needs.
    
    \subsubsection{Selected Hardware} \label{selected-hardware}
    
        We purchased, evaluated, and deployed three CAN bus interfaces that provided reliable data collection for our research needs:
        
        \paragraph{Korlan USB2CAN} This adapter connects a computer's USB port to a vehicle's OBD-II port, enabling direct CAN bus monitoring---and message injection. The device functions as a network node within the CAN infrastructure, leveraging the broadcast nature of the CAN protocol to monitor all network traffic. Key specifications include:
        
        \begin{itemize}
            \item Direct CAN bus access eliminates diagnostic query limitations
            \item Wired CAN-to-USB interface reduces latency and data loss, improving reliability
            \item Bidirectional CAN frame transmission and reception supports data collection and message injection
            \item Drivers, documentation, and example code help practitioners to get started
            \item Cost: 69 USD (as of this writing) \cite{8devices-obd2-version}
        \end{itemize}

        The Korlan USB2CAN adapter represents the most cost-effective solution maintaining acceptable data quality comparable to the significantly more expensive alternatives described below.

        \begin{figure}[htb!]
        \caption{Our two data collection strategies.}
            \begin{subfigure}[t]{0.49\textwidth}
            \centering
            \caption{A Linux laptop connects to the CAN bus via a CAN-to-USB cable. Data is collected using SocketCAN's \texttt{candump} command.}
            \includegraphics[width=1.0\textwidth]{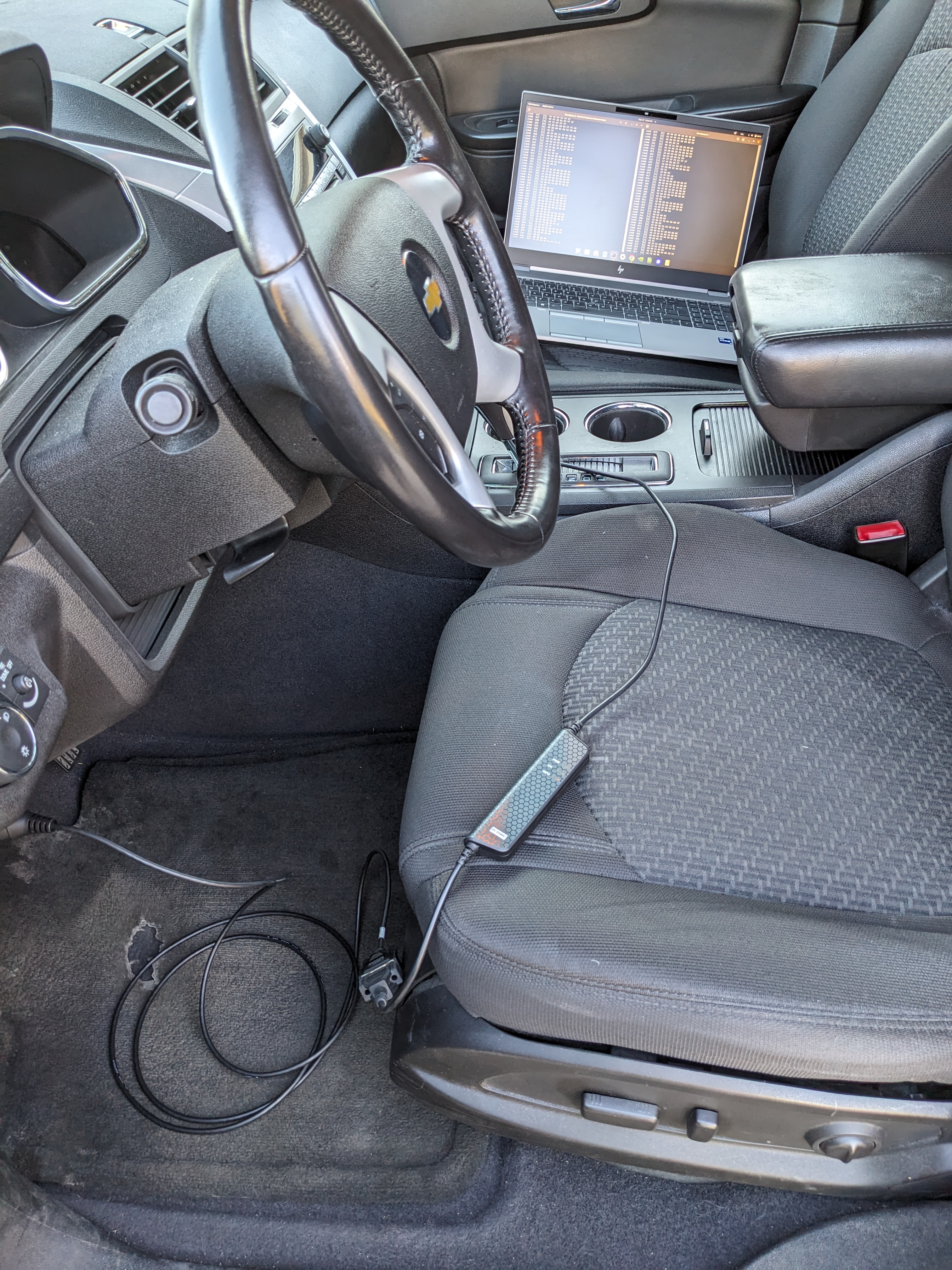}
            \label{setup-with-laptop}
            \end{subfigure}
            \hfill
            \begin{subfigure}[t]{0.49\textwidth}
            \centering
            \caption{A standalone data collection device---CANEdge2---connects to the CAN bus via a CAN-to-DB9 cable.}
            \includegraphics[width=1.0\textwidth]{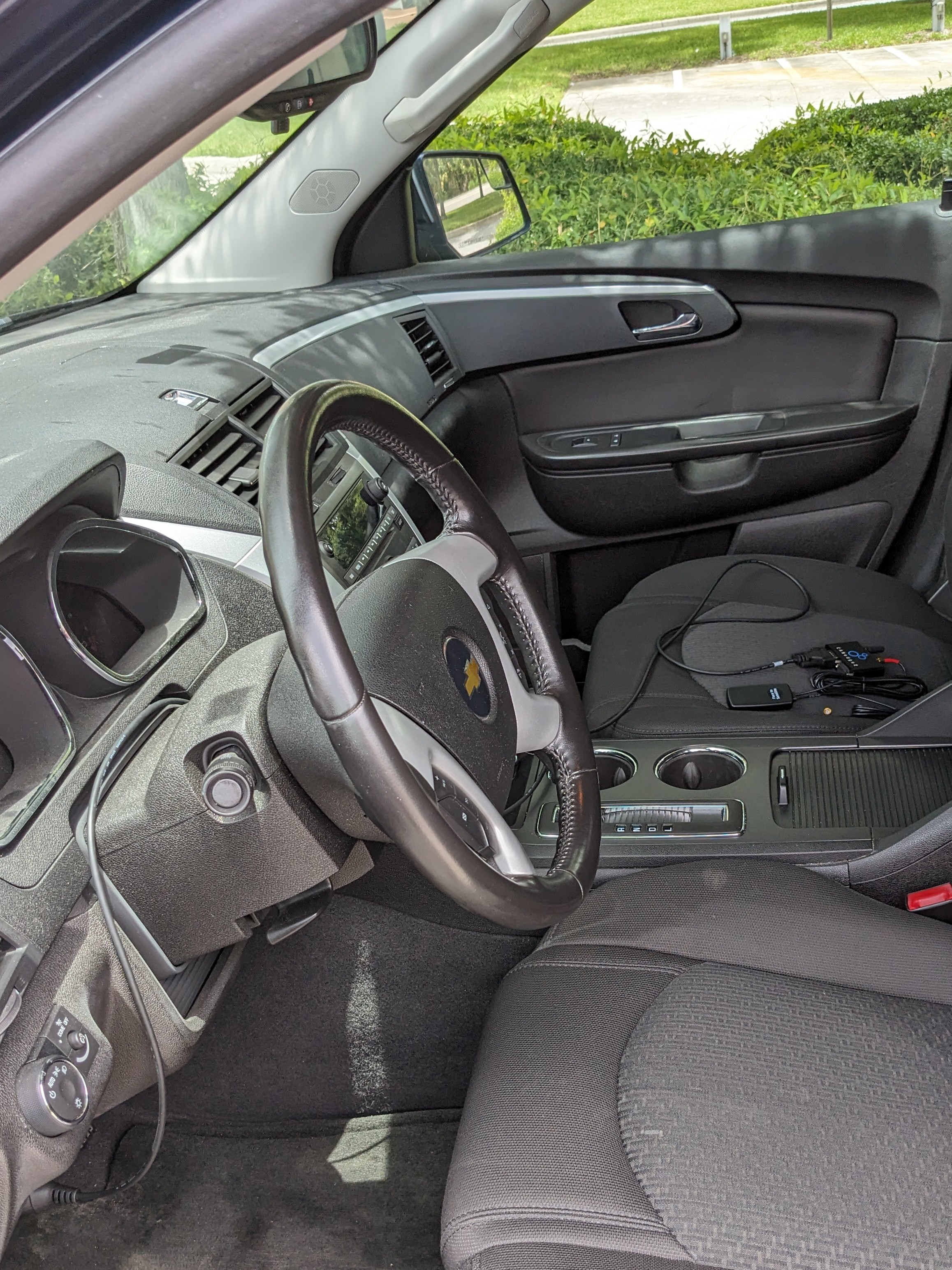}
            \label{setup-with-canedge}
            \end{subfigure}
        \end{figure}

        \paragraph{Kvaser Hybrid CAN-LIN} This professional-grade adapter provides advanced CAN bus monitoring---and message injection---capabilities with additional features for specialized applications. The device requires a separate OBD-II to DB9 cable for vehicle connectivity. Key specifications include:
        
        \begin{itemize}
            \item Direct, wired CAN bus access
            \item Bidirectional CAN frame transmission and reception
            \item Local Interconnect Network (LIN) channel (unused in our research)
            \item Proprietary drivers, professional software suite, development kits \& tools, and extensive documentation
            \item Cost: 501 USD \cite{kvaser-can-lin}
        \end{itemize}
        
        This solution provides enhanced functionality for practitioners who need LIN protocol access or seek specialized drivers, software, or tools. The Kvaser Hybrid CAN-LIN adapter is shown in Figure \ref{setup-with-laptop}.
        
        \paragraph{CSS Electronics CANEdge2} This autonomous data logger facilitates high-quality data collection over extended monitoring periods. The device stores CAN data on a removable SD card and uploads data to cloud storage services when connected to Wi-Fi networks. The device requires a separate OBD-II to DB9 cable for vehicle connectivity. Key specifications include:
        
        \begin{itemize}
            \item Autonomous operation; no computer needed
            \item SD card local storage with cloud synchronization
            \item AWS S3 compatibility (used in our research)
            \item Direct, wired CAN bus access
            \item Bidirectional CAN frame transmission and reception
            \item Local Interconnect Network (LIN) channel (unused in our research)
            \item Free and open-source software and APIs included
            \item Cost: 480 EUR \cite{canedge2}
        \end{itemize}
        
        While the CANEdge2 provides convenient autonomous operation, it reduces practitioner control. Saving each trip as a separate CAN trace can be challenging, as can labeling to keep track of, e.g., who was driving and where. Practitioners should plan accordingly when collecting data from multiple drivers. The CSS Electronics CANEdge2 data logger is shown in Figure \ref{setup-with-canedge}.

\subsection{Dataset Composition}

    KCID addresses critical limitations in existing driver fingerprinting datasets by providing raw CAN bus data from 16 drivers across four vehicles, representing diverse demographic groups, driving styles, and vehicle types. This section details the vehicles, drivers, routes, and temporal characteristics of the dataset.

    \subsubsection{Vehicles}
    
        Data collection encompassed four vehicles representing different manufacturers, model years, and vehicle classifications. Table \ref{tab:vehicles} summarizes vehicle characteristics and the number of volunteer drivers per vehicle.
        
        \begin{table*}[htb!]
        \small
        \centering
            \begin{threeparttable}
            \caption{Vehicles included in our study} \label{tab:vehicles}
            
            \begin{tabular}{|p{3cm}|p{1cm}|p{4cm}|p{2cm}|p{2cm}|} \hline
                \rowcolor{LightSteelBlue3!80}
                \textbf{Vehicle Model} & \textbf{Year} & \textbf{Vehicle Type} & \textbf{Drive Type} & \textbf{\# of Drivers\tnote{1}} \\ \hline \hline
                
                \rowcolor{LightSteelBlue3!40}
                Chevrolet Traverse & 2011 & 5-door full-size SUV crossover & AWD & 8\tnote{2} \\ \hline
                
                \rowcolor{LightSteelBlue3!10}
                Ford Focus & 2017 & 5-door compact station wagon & FWD & 4 \\ \hline
                
                \rowcolor{LightSteelBlue3!40}
                Subaru Forester & 2017 & 5-door compact SUV crossover & AWD & 6\tnote{3} \\ \hline
                
                \rowcolor{LightSteelBlue3!10}
                Honda CR-V Touring & 2022 & 5-door compact SUV crossover & AWD & 1 \\ \hline
            \end{tabular}
            \footnotesize
            \begin{tablenotes}
                \item[1] The ``\# of Drivers'' column includes volunteer drivers whose data was captured in single-driver traces, where we know who was driving at all times. We exclude volunteer drivers whose data is only available in mixed traces because we do not know when each specific driver was actually operating the vehicle.
                \item[2] 8 unique drivers in single-driver traces; 1 additional driver in a mixed trace
                \item[3] 6 unique drivers in single-driver traces; 3 additional drivers in mixed traces
            \end{tablenotes}
            \end{threeparttable}
        \end{table*}

        Our vehicle selection was primarily determined by availability---vehicle owners who were willing to allow data collection from their vehicles or lend their vehicles for data collection purposes. We were fortunate to include four different manufacturers---Chevrolet, Ford, Subaru, and Honda---providing cross-manufacturer representation in the dataset.

        The availability and utility of CAN bus data varies depending on the vehicle. The Traverse, Focus, and Forester provide substantial CAN data; we observed approximately 50-70 unique arbitration IDs during our driving sessions. We have concentrated our reverse engineering efforts on these three vehicles, especially the Traverse, which we leverage for our proof of concept implementation described in Section \ref{anti-theft-framework-and-proof-of-concept}. For the Traverse and the Forester, our reverse engineering efforts---augmented by publicly available reverse engineering data---have provided insights into the CAN data and confirmed that there are ample behavioral signals suitable for driver fingerprinting and authentication applications. The Focus presents a greater analytical challenge due to limited reverse engineering data and documentation on our specific model year; nevertheless, we observed distinguishable patterns in the CAN traffic corresponding to different vehicle states and driver interactions with various vehicle features. The Honda CR-V provides markedly different CAN traffic, with a mere six unique arbitration IDs observed, many containing repetitive data field content. While some data fields demonstrate variation that may prove useful, substantial reverse engineering effort would be required to fully characterize their potential for driver fingerprinting and authentication applications.

        \begin{figure}[htb!]
        \caption{The on-board diagnostics II (OBD-II) ports of two vehicles used in this study. We leveraged the OBD-II ports to collect data for our driver authentication dataset. A data cable (connected to either a laptop or a standalone data collection device) would be plugged into the port to facilitate data collection.}
            \begin{subfigure}[t]{0.49\textwidth}
            \centering
            \caption{The OBD-II port of a 2011 Chevrolet Traverse.}
            \includegraphics[width=1.0\textwidth]{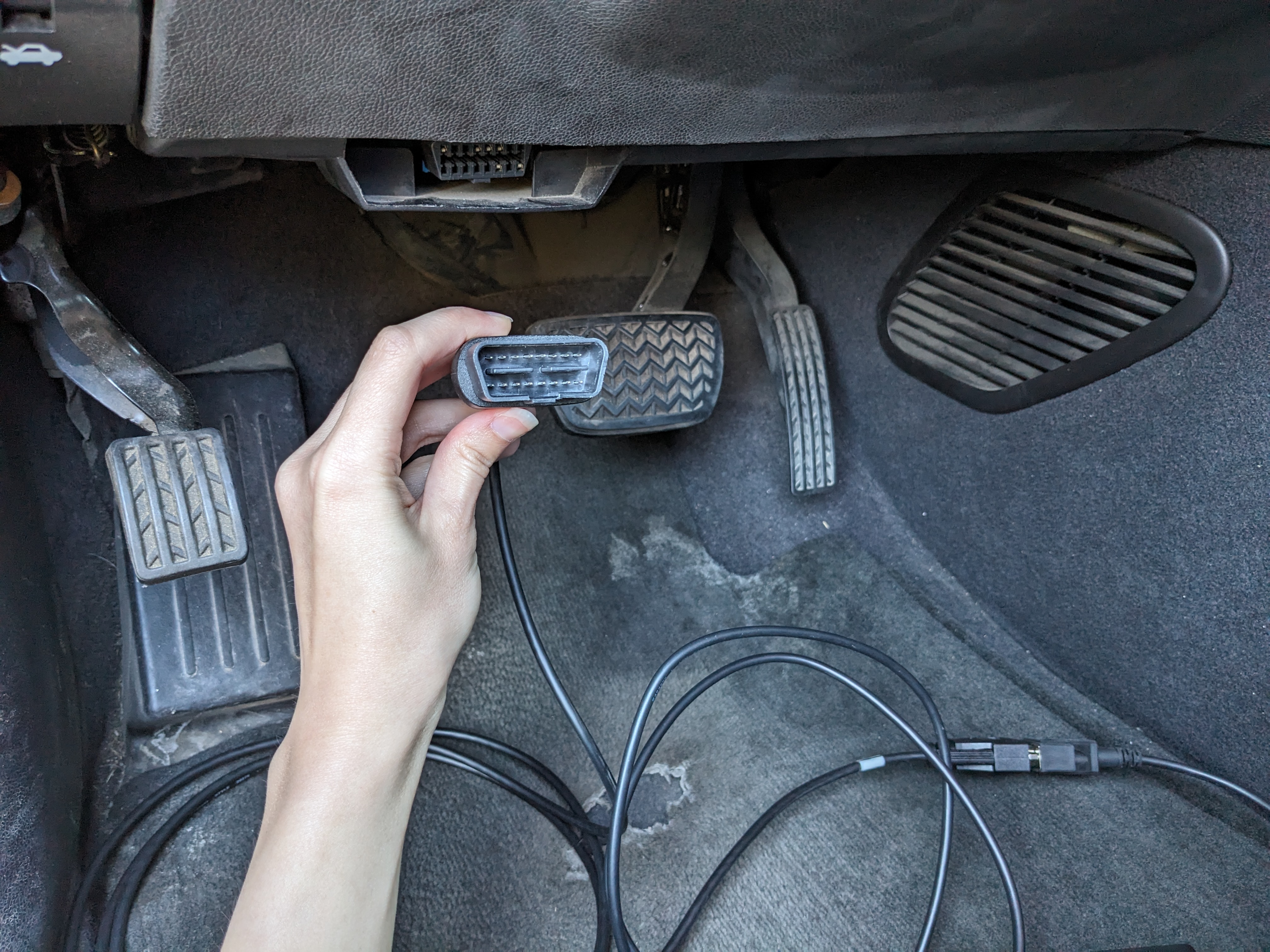}
            \label{traverse-obd}
            \end{subfigure}
            \hfill
            \begin{subfigure}[t]{0.49\textwidth}
            \centering
            \caption{The OBD-II port of a 2017 Subaru Forester.}
            \includegraphics[width=1.0\textwidth]{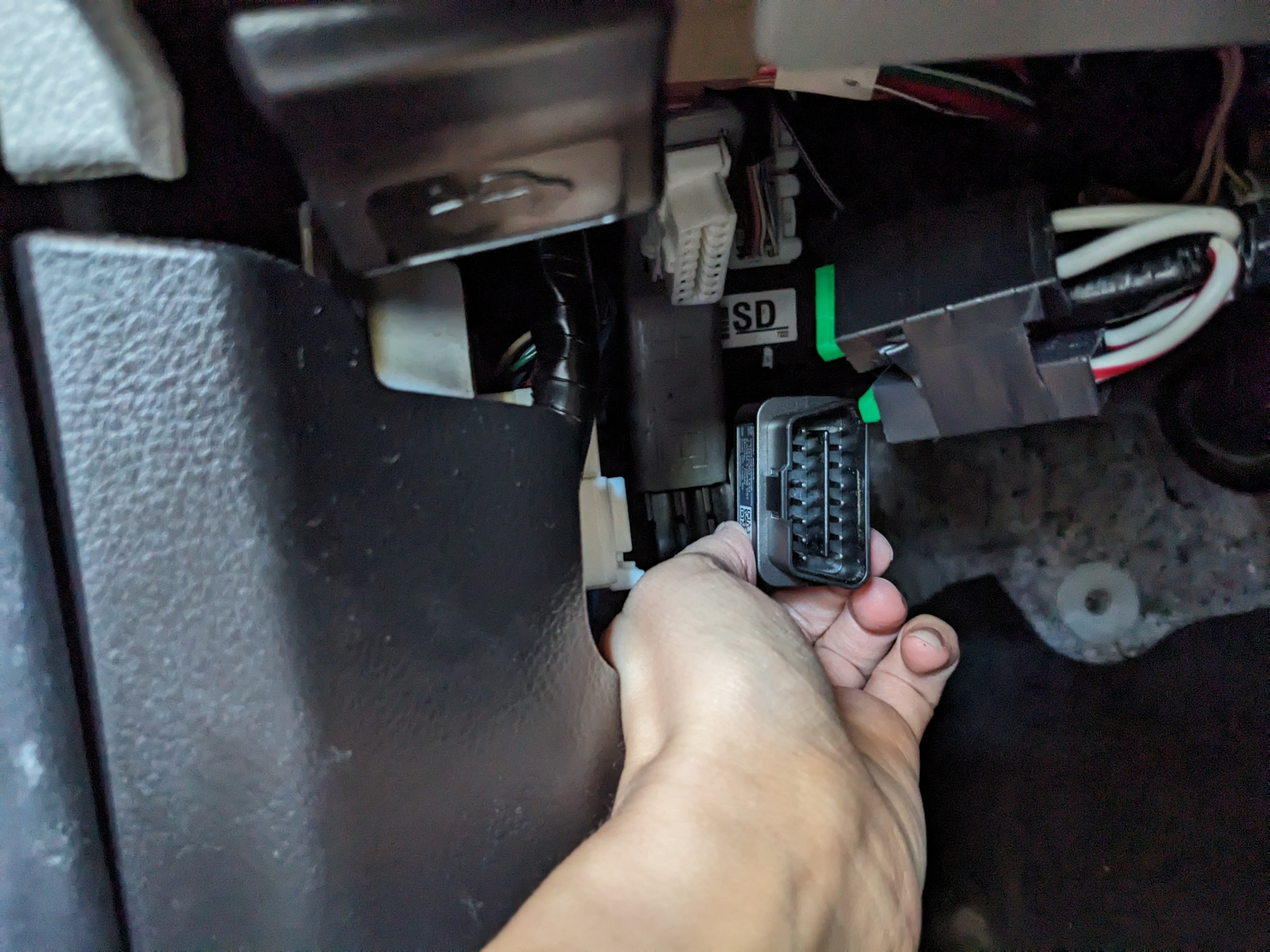}
            \label{forester-obd}
            \end{subfigure}
        \end{figure}

        Figures \ref{traverse-obd} and \ref{forester-obd} show the OBD-II ports of a 2011 Chevrolet Traverse and a 2017 Subaru Forester. We connect our data collection hardware (described in Section \ref{selected-hardware}) to these ports to access the vehicles' CAN buses. U.S. federal law requires the OBD-II port to be within arm's reach of the driver's seat and accessible without tools \cite{obdii-law}. Similar legislation exists in the European Union through the European on-board diagnostics (EOBD) regulations \cite{eobd-law}. The U.S. requirement has been in effect since 1996 \cite{history}, and the EU requirement came into force in 2001 \cite{eobd-court-ruling}. A recent ruling by the Court of Justice of the European Union suggests this data will remain accessible for the foreseeable future \cite{eobd-court-ruling}.

        While OBD-II and EOBD diagnostic data follow standardized protocols, raw CAN bus traffic is manufacturer- and model-specific, so reverse engineering---on a per-vehicle basis---may be necessary. Fortunately, several automated and semi-automated reverse engineering methods are available: \cite{extracting-signals,can-reverse-engineering,recan-dataset,classification-ids}.
        
    \subsubsection{Drivers}

        KCID includes data from 16 drivers representing diverse demographic categories, including multiple age groups, both sexes, two nationalities, and various levels of driving experience. This demographic diversity addresses a critical limitation identified in Section \ref{research-gaps-and-dataset-limitations}: the absence of demographic information in existing driver fingerprinting datasets, which prevents researchers from evaluating driver authentication system performance across different user populations. Table \ref{tab:dataset_stats} showcases the collected CAN messages on a per driver, per vehicle basis, with drivers organized by demographic category.

        \begin{table}[htb!]
            \centering
            \begin{threeparttable}
            \caption{Distribution of collected CAN messages per driver and per vehicle.\tnote{1}} \label{tab:dataset_stats}
            \begin{tabular}{>{\raggedright\arraybackslash}p{3cm}>{\raggedleft\arraybackslash}p{2cm}>{\raggedleft\arraybackslash}p{2cm}>{\raggedleft\arraybackslash}p{2cm}>{\raggedleft\arraybackslash}p{2cm}>{\raggedleft\arraybackslash}p{2cm}}
                \toprule
                Driver\tnote{2} & \multicolumn{1}{>{\raggedright\arraybackslash}p{2cm}}{2011 Chevrolet Traverse} & \multicolumn{1}{>{\raggedright\arraybackslash}p{2cm}}{2017 Ford Focus} & \multicolumn{1}{>{\raggedright\arraybackslash}p{2cm}}{2017 Subaru Forester} & \multicolumn{1}{>{\raggedright\arraybackslash}p{2cm}}{2022 Honda CR-V Touring} & \multicolumn{1}{>{\raggedright\arraybackslash}p{2cm}}{Total}  \\
                \midrule
                \texttt{female-all-ages-1} & 0 & 0 & 3,077,765 & 0 & 3,077,765\\
                \texttt{female-all-ages-2} & 145,025,032 & 0 & 77,463,024 & 0 & 222,488,056\\
                \texttt{female-all-ages-4} & 0 & 0 & 0 & 7,845,042 & 7,845,042 \\
                \texttt{female-all-ages-5} & 343,044,443 & 19,286,594 & 21,250,993 & 0 & 383,582,030\\ 
                \midrule 
                \texttt{male-under30-1} & 2,086,138 & 0 & 0 & 0 & 2,086,138\\
                \texttt{male-under30-2} & 42,446,559 & 0 & 0 & 0 & 42,446,559\\
                \texttt{male-under30-3} & 55,683,325 & 0 & 0 & 0 & 55,683,325\\
                \texttt{male-under30-4} & 0 & 8,708,825 & 0 & 0 & 8,708,825\\
                \midrule
                \texttt{male-30-55-1} & 0 & 9,156,269 & 0 & 0 & 9,156,269 \\
                \texttt{male-30-55-2} & 0 & 0 & 8,271,882 & 0 & 8,271,882 \\
                \texttt{male-30-55-3} & 3,454,936 & 0 & 67,715,089 & 0 & 71,170,025\\
                \texttt{male-30-55-4} & 42,318,998 & 0 & 0 & 0 & 42,318,998\\
                \midrule 
                \texttt{male-over55-1} & 0 & 0 & 22,352,130 & 0 & 22,352,130 \\
                \texttt{male-over55-2} & 109,130,957 & 14,858,114 & 0 & 0 & 123,989,071\\
                \midrule
                Total & 743,190,388 & 52,009,802 & 200,130,883 & 7,845,042 & 1,003,176,115
            \end{tabular}
            \footnotesize
            \begin{tablenotes}
                \item[1] We include volunteer drivers whose data was captured in single-driver traces, where we know who was driving at all times. We exclude volunteer drivers whose data is only available in mixed traces because we do not know when each specific driver was actually operating the vehicle. Drivers \texttt{female-all-ages-3} and \texttt{male-over55-3} are not included in the table.
                \item[2] Due to limited sample size, female drivers are not subdivided by age category. However, female participants represent all three age groups (under 30, 30-55, and over 55), ensuring demographic coverage across the age spectrum for both sexes.
            \end{tablenotes}
            \end{threeparttable}
        \end{table}
        
        KCID represents substantial data collection efforts spanning more than a year. As shown in Table \ref{tab:dataset_stats}, even when excluding mixed-driver traces (driving sessions with multiple drivers), single-driver traces alone comprise over one billion CAN messages distributed across 14 drivers. This volume substantially exceeds that of many existing open-access driver fingerprinting datasets, providing researchers with the data scale necessary for training robust driver authentication models and conducting rigorous testing and validation.
        
        \textbf{Demographics.} The dataset includes 16 drivers distributed across different demographic categories. Male drivers are divided into three age groups: 4 drivers under 30 years of age (\texttt{male-under30-1} through \texttt{male-under30-4}), 4 drivers between 30-55 years of age (\texttt{male-30-55-1} through \texttt{male-30-55-4}), and 3 drivers over 55 years of age (\texttt{male-over55-1} through \texttt{male-over55-3}). The dataset includes 5 female drivers (\texttt{female-all-ages-1} through \texttt{female-all-ages-5}). While privacy considerations precluded subdividing female drivers into comparable age categories due to limited sample size, our female participants represent all three age groups, ensuring demographic coverage across the age spectrum for both sexes.
        
        Our participants were recruited through the authors' professional and personal networks, with several authors contributing their own driving data (collected via autonomous CANEdge2 logging, a data collection laptop configured to collect data autonomously before departure, or by a co-author operating the data collection laptop). While volunteer-based recruitment does not guarantee perfect population representativeness, we prioritized capturing the demographic dimensions most salient for driver authentication research—specifically, sex and age. These factors demonstrably influence driving behavior: Ahmadi-Assalemi et al. \cite{uk-driving-dataset} identified significant sex-based differences in driving patterns that affected driver authentication system performance, and insurance industry data establishes strong correlations between age, sex, and driving risk profiles \cite{fatality-facts,age-gender-forbes} (see Section \ref{research-gaps-and-dataset-limitations}).
        
        Our participant pool exhibits diverse driving frequencies, ranging from daily commuters who drive regularly for work or personal activities to occasional drivers who primarily rely on alternative transportation modes. Geographically, our dataset encompasses both U.S. and Danish drivers, with some participants driving in both countries. This international composition provides cross-cultural representation, capturing variation in driving behaviors shaped by different traffic regulations, road infrastructure, and cultural driving norms. Future work includes expanding the participant pool substantially to enable more granular demographic disclosure without unduly impacting volunteer drivers' privacy.

    \subsubsection{Devices}
    
        \begin{table}[htb!]
            \centering
            \begin{threeparttable}
            \caption{Distribution of collected CAN messages per data collection device.\tnote{1}} \label{tab:total_device}
            \begin{tabular}{>{\raggedleft\arraybackslash}p{2.5cm}>{\raggedleft\arraybackslash}p{2.5cm}>{\raggedleft\arraybackslash}p{2.0cm}>{\raggedleft\arraybackslash}p{2.5cm}}
                \toprule
                 CSS Electronics CANEdge2 & Kvaser Hybrid CAN-LIN & Korlan USB2CAN & Total \\
                 \midrule
                 852,868,766 & 98,297,547 & 52,009,802 & 1,003,176,115
            \end{tabular}
            \footnotesize
            \begin{tablenotes}
                \item[1] For consistency, we include CAN messages from single-driver traces; we exclude CAN messages from mixed traces.
            \end{tablenotes}
            \end{threeparttable}
        \end{table}
        
        As described in Section \ref{selected-hardware}, data collection was performed using three different CAN interface devices: Korlan USB2CAN, Kvaser Hybrid CAN-LIN, and CSS Electronics CANEdge2. Table \ref{tab:total_device} presents the total number of CAN messages collected by each device across all single-driver traces.
        
    \subsubsection{Locations}
    
        Data collection occurred across multiple locations in two countries: Denmark and the United States (mainly Florida and Nebraska). U.S. data collection included both local and regional driving within these states and long-distance interstate trips between Florida and Nebraska, with one route including a stopover in Tennessee. This geographic diversity encompasses driving conditions, road types, traffic patterns, driving norms, traffic laws, and environmental factors, enhancing the dataset's representativeness for real-world authentication applications.

    \subsubsection{Route Types}
    
        KCID includes two types of routes to serve different research needs:
        
        \textbf{Daily Driving Routes.} Most of KCID captures natural, unscripted driving during volunteer drivers' normal daily activities. This data reflects authentic driving patterns---the route choices, behavioral variations, and conditions that characterize each driver's normal vehicle use. Daily driving data includes urban and residential streets, highway and toll-road/expressway travel, and parking maneuvers.
        
        \textbf{Fixed Routes.} To complement the daily driving data and enable controlled comparisons, KCID includes a specialized fixed routes experiment (see Section \ref{fixed-routes-experiment}). Six drivers (\texttt{male-30-55-3}, \texttt{male-30-55-4}, \texttt{male-over55-1}, \texttt{female-all-ages-1}, \texttt{female-all-ages-2}, and \texttt{female-all-ages-5}) drove two vehicles (2011 Chevrolet Traverse and 2017 Subaru Forester) along predetermined routes in Florida. Fixed routes minimize route-dependent noise---such as the differences between short local trips and long highway journeys---thereby isolating driver-specific behavioral patterns. This controlled approach is particularly valuable for initial model training and feature engineering, though daily driving data remains essential for developing robust authentication systems suitable for real-world deployment.

    \subsubsection{File Formats}

        To maximize accessibility across different research workflows and software ecosystems, KCID provides data in three formats:

        \begin{enumerate}
            \item \textbf{.mf4 (MDF4)} Binary format standardized by the Association for Standardization of Automation (ASAM), offering compact file sizes and native compatibility with automotive-specific analysis tools. This is the native output format from the CSS Electronics CANEdge2. Only CAN traces collected by CANEdge2 are available in this format.
            
            \item \textbf{.log} Text-based format compatible with Linux SocketCAN can-utils, enabling straightforward integration with Python's \texttt{python-can} library. Researchers can replay log files to virtual CAN interfaces, enabling them to process and analyze data as though receiving live CAN bus streams.
            
            \item \textbf{.csv} Text-based comma-separated values format that can be easily processed by Python's pandas library as well as popular Python-based machine learning frameworks (scikit-learn, Keras, TensorFlow, PyTorch).
        \end{enumerate}

        This multi-format approach allows researchers to select the most appropriate format for their specific methodologies and tools without needing to do a format conversion.

    \subsubsection{Specialized Experiments} \label{specialized-experiments}
    
        Beyond general driving data, KCID includes five specialized data collection experiments designed to support specific research applications:
        
        \begin{enumerate}
            \item \textbf{Fixed Routes Experiment:} Volunteer drivers drove predetermined, mappable routes, eliminating route-based noise in the driving data and enabling greater differentiation between drivers (see Section \ref{fixed-routes-experiment}).
            \item \textbf{OBD Requests and Responses Experiment:} One driver drove while CANEdge2 systematically queried 16 OBD-II PIDs, providing diagnostic data to aid in reverse engineering and driver fingerprinting efforts.
            \item \textbf{Tire Pressure Experiment:} One driver drove multiple short loops while we collected data on the effects of tire pressure on CAN bus signals.
            \item \textbf{Driving Modes and Features Experiment:} One driver drove in a low-traffic area while engaging various vehicle functions (gear selection, turn signals, etc.), allowing us to collect function- and feature-specific CAN traces for future reverse engineering efforts.
            \item \textbf{Stationary Vehicles Experiment:} We collected data from two very new, very modern vehicles (2024 Chevrolet Malibu, 2025 Toyota Corolla) while they sat stationary with the engines running, facilitating a comparison between the CAN traffic of older vehicles---such as the 2011 Chevrolet Traverse included in our dataset---and newer vehicles.
        \end{enumerate}
        
        Each specialized experiment includes detailed documentation in a dedicated \texttt{README.md} file within its respective dataset subdirectory, covering experimental protocols and data organization.

    \subsubsection{Fixed Routes Experiment} \label{fixed-routes-experiment}
    
        This experiment systematically collected CAN traces over predetermined routes to enable controlled comparison of driver-specific behavioral patterns. By standardizing route characteristics, this approach minimizes confounding variables arising from differences in everyday driving habits (e.g., short local trips versus long-distance highway travel), facilitating more precise isolation of individual driving signatures. While Section \ref{research-gaps-and-dataset-limitations} identified unrealistic fixed-route designs as a limitation in existing datasets, controlled route data remains valuable for specific research applications, particularly feature engineering, model training, and controlled evaluation of driver authentication algorithms.
        
        \begin{figure*}[htb!]
            \centering
            \includegraphics[width=1.0\textwidth]{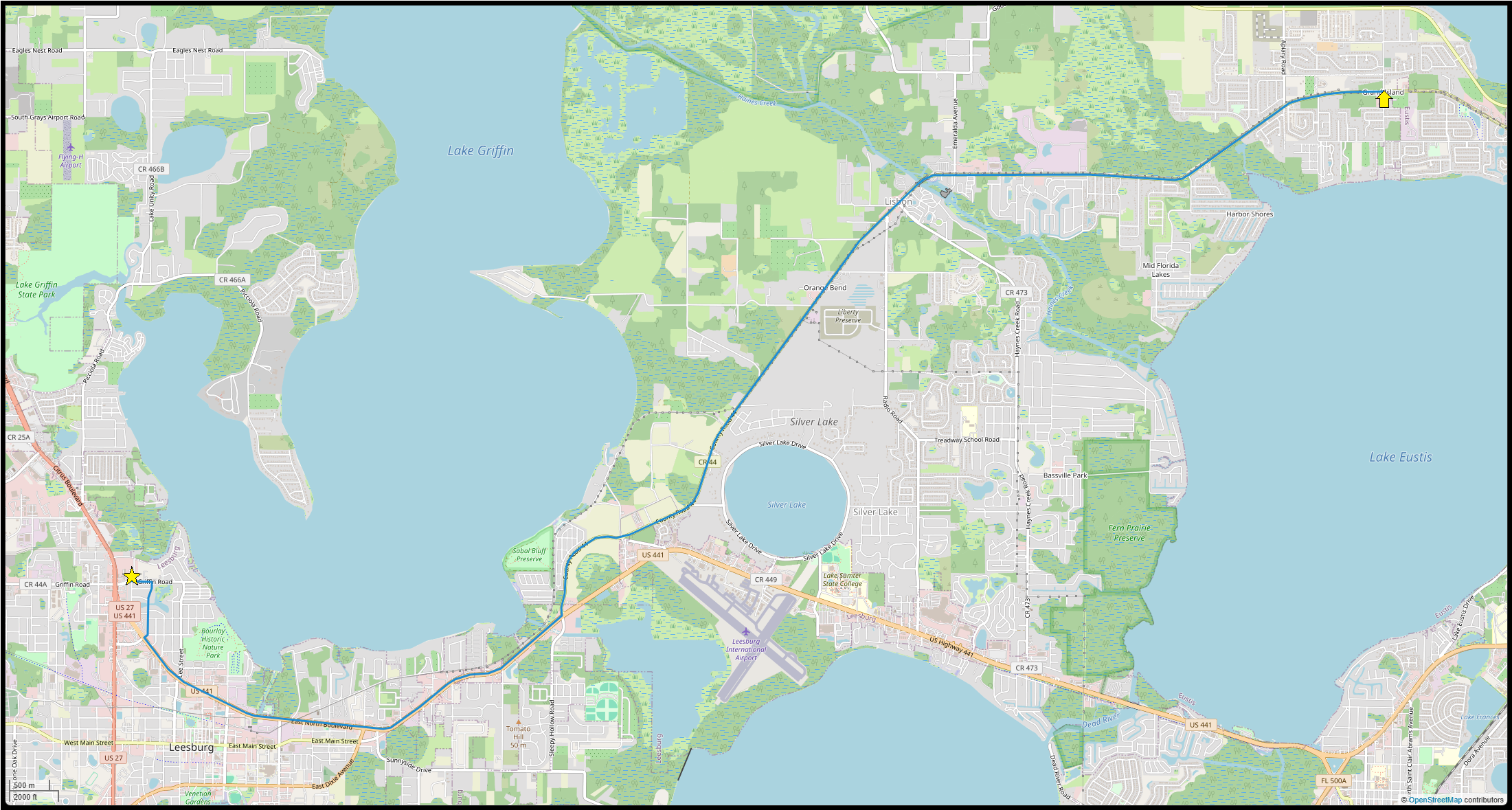}
            \caption{Fixed route from Grand Island, FL, to Bethany Lutheran Church in Leesburg, FL. \newline \textbf{© OpenStreetMap contributors.} Available under the Open Data Commons Open Database License (ODbL). \newline See \href{https://www.openstreetmap.org/copyright}{\nolinkurl{openstreetmap.org/copyright}}.}
            \label{fixed-route}
        \end{figure*}
        
        \textbf{Primary Route: Grand Island to Leesburg.} The primary fixed route connects Grand Island to Bethany Lutheran Church in Leesburg, Florida---approximately 12.6 miles (20 kilometers) with typical travel times between 15 and 25 minutes. Figure \ref{fixed-route} shows the route overlaid on OpenStreetMap. The route is also available on Google Maps: \href{https://maps.app.goo.gl/LEiAnmhP7J4YPze58}{\nolinkurl{maps.app.goo.gl/LEiAnmhP7J4YPze58}}
        
        We selected this route because multiple volunteer drivers regularly traveled it regardless of data collection activities, ensuring naturalistic driving behavior while providing route consistency. To protect privacy, the origin is specified as the center of Grand Island (intersection of S Fish Camp Road and County Road 44) using the address of a gas station, rather than the actual private residence. The destination is a semi-public location shown at its exact address.
        
        Six drivers contributed data over this route, with systematic variation in departure times capturing different traffic conditions. Sunday morning departures (approximately 9:50-10:00 AM) encountered minimal traffic, while Wednesday and Thursday evening departures (approximately 4:50-5:00 PM) occurred during rush hour with substantially increased congestion. This temporal variation enables researchers to assess how traffic conditions influence driving patterns and authentication performance.
        
        \textbf{Additional Routes.} Beyond the primary route, the dataset includes additional fixed routes connecting Grand Island to various destinations: HomeGoods in Clermont, Orlando International Airport (MCE), Orlando Sanford International Airport (SFB), and the University of Central Florida. These routes encompass diverse road types and driving contexts, with some routes captured using both toll and non-toll alternatives, providing additional opportunities to compare and analyze driving behavior under different conditions. Route selection prioritized locations that volunteer drivers regularly visit, optimizing data collection within time and budget constraints while ensuring naturalistic driving behavior.
        
        \textbf{Notes.} For CANEdge2-collected traces, individual trips may span multiple files due to automatic file segmentation; all files from a single trip are grouped within one folder to maintain trace continuity.
        
        Further documentation is available in the \texttt{README.md} file in KCID's \texttt{FIXED-ROUTES-EXPERIMENT} subdirectory.
        
\subsection{Privacy Considerations} \label{privacy-considerations}

    All volunteer drivers were informed of the potential privacy implications and provided informed consent for data collection. Given the limited sample size, drivers acknowledged that external data sources (e.g., people search websites) could potentially be used to identify them if they were named in the acknowledgments. Accordingly, some drivers opted not to be named for privacy reasons, while others wished to be acknowledged with full awareness of the privacy implications.
    
    The dataset reveals driving patterns including acceleration, speed, turning, and braking habits. Extracting this information requires non-trivial reverse engineering effort, and the data lacks environmental context such as traffic control devices, speed limits, and pedestrian or vehicle interactions. A sudden braking event, for example, could reflect either driving style or an appropriate response to avoid an accident---the data alone cannot distinguish between these scenarios. The fixed routes portion provides somewhat more information because researchers could reference speed limits and traffic control systems along those routes. However, variable traffic conditions and our omission of the exact start and/or end points prevent researchers from precisely pinpointing the vehicle location at any given time. This limitation constrains potential misuse while preserving sufficient information for driver authentication research.
    
    No driver is uniquely identified within their demographic category---the smallest category contains three drivers. More importantly, the data itself cannot reasonably facilitate driver de-anonymization without a comprehensive database of drivers and their driving patterns. Anyone who has access to such a database would not need our dataset to de-anonymize our drivers.
    
    For example, many automotive manufacturers collect comprehensive data about their customers, their vehicles, and anyone who drives or rides in them \cite{cars-and-privacy,cars-privacy-not-included,smart-vehicles-in-smart-cities}. They collect far more data than we do, including personal, sensitive data that we do not collect. They would neither need nor benefit from using our dataset to identify our volunteer drivers. In other words, the only parties who could potentially identify unnamed drivers are those who already possess comprehensive data about those specific individuals, making our dataset's contribution to de-anonymization risk negligible.

\subsection{Data Availability and Access}

The Kidmose CANid Dataset (KCID) is publicly available through DTU Data at \doi{10.11583/DTU.30483005} \cite{kcid-dtu-data}. The dataset includes all CAN trace files in multiple formats (.log, .csv, and .mf4) and comprehensive documentation as described in Section \ref{kcid}. Dataset documentation will be updated to reflect any revisions made during the peer review process and upon publication to include complete citation information.

\section{Applications} \label{applications}

The Kidmose CANid Dataset supports diverse applications unified by a common foundation: analyzing behavioral patterns in CAN bus data. Driver authentication---distinguishing authorized from unauthorized users to prevent theft---represents the primary application explored throughout this paper, including our anti-theft framework in Section \ref{anti-theft-framework-and-proof-of-concept}. The same raw CAN traffic and driving behavior patterns that enable driver authentication also support insurance risk profiling, safety monitoring for young and fleet drivers, detection of impaired or reckless driving, and mechanical anomaly identification. This section examines these applications, demonstrating how KCID's comprehensive data facilitates research across automotive cybersecurity, safety, and diagnostic domains.

\subsection{Driver Authentication} \label{machine-learning}

    Driver authentication---distinguishing authorized drivers from unauthorized individuals---constitutes KCID's primary application and motivates the anti-theft framework demonstrated in Section \ref{anti-theft-framework-and-proof-of-concept}. As discussed in Section \ref{driver-fingerprinting-and-authentication}, existing driver authentication research has consistently employed machine learning approaches to extract behavioral signatures from vehicle data \cite{driver-fingerprinting,hcrl-driving-dataset,auto-theft-detection-clustering,this-car-is-mine-dataset,this-car-is-mine-dataset-escar,uk-driving-dataset,theft-detection-can-bus,extracting-signals}. KCID's raw CAN data, demographic diversity, authentic driving conditions, and specialized experiments enable researchers to develop and evaluate driver authentication systems.
    
    Fundamentally, driver authentication constitutes a binary classification problem similar to intrusion detection: does CAN traffic originate from an ``authorized driver'' or an ``unauthorized driver?'' Many existing automotive intrusion detection systems can be adapted for driver authentication applications \cite{can-fp}. Both supervised methods (using labeled authorized and unauthorized driver data) and unsupervised methods (establishing authorized baselines and detecting deviations) can be applied to driver authentication in general and the Kidmose CANid Dataset in particular.
    
    Techniques applicable to KCID for driver fingerprinting and authentication include:
    
    \begin{enumerate}
        \item \textbf{Feature Engineering and Dimensionality Reduction.} Raw CAN features (timestamp, arbitration ID, data field) can be transformed and enhanced for authentication purposes. Temporal features (inter-message time deltas), data field decompositions (individual bytes, decoded signals), and dimensionality reduction techniques (PCA, t-SNE, UMAP) enable extraction of behavioral signatures while reducing noise \cite{classification-ids,all-learning-survey,pca-data-exploration,anomaly-detection-statistics}.
    
        \item \textbf{Traditional Machine Learning: Classifiers.} Various classifiers have proven effective for driver authentication, including decision trees, random forests, support vector machines, naïve Bayes, and $k$-nearest neighbors \cite{driver-fingerprinting,hcrl-driving-dataset}. KCID enables comparative evaluation of these approaches on raw CAN data across multiple vehicles and demographic groups.
    
        \item \textbf{Traditional Machine Learning: Clustering.} Clustering algorithms (e.g., $k$-means) enable driver authentication without requiring labeled ``unauthorized driver'' data. Instead, clustering-based approaches establish ``authorized driver'' behavioral profiles and detect unauthorized access through cluster deviation analysis \cite{auto-theft-detection-clustering}. This approach is particularly practical since researchers cannot realistically obtain training data from actual vehicle thieves.
    
        \item \textbf{Deep Learning.} Artificial neural networks (ANNs), including autoencoders and long short-term memory (LSTM), have demonstrated their effectiveness for driver fingerprinting and authentication applications. Autoencoders trained on authorized drivers' data achieve minimal reconstruction error for legitimate users but exhibit elevated error for unauthorized drivers \cite{autoencoder-anomaly-detection}. LSTM networks excel at modeling temporal dependencies in CAN message sequences, learning authorized drivers' patterns and detecting deviations \cite{lstm-4,theft-detection-can-bus,all-learning-survey}.
    
        \item \textbf{Time Series Analysis.} CAN data's inherent temporal structure facilitates time series analysis. Lag features incorporating previous or successive CAN messages as additional dimensions can capture temporal dependencies \cite{scikit-lagged-features,deep-learning-time-series-survey,anomaly-detection-time-series,deep-learning-time-series}.
    \end{enumerate}
    
\subsection{Driver Profiling for Insurance and Safety}

\begin{figure}[htb!]
        \caption{Real-time driver monitoring by State Farm car insurance, showing (a) overall safety scores across multiple categories and (b) detailed driving events from a single trip.}
            \begin{subfigure}[t]{0.49\textwidth}
            \centering
            \caption{Driving event summary}
            \includegraphics[width=0.7\textwidth]{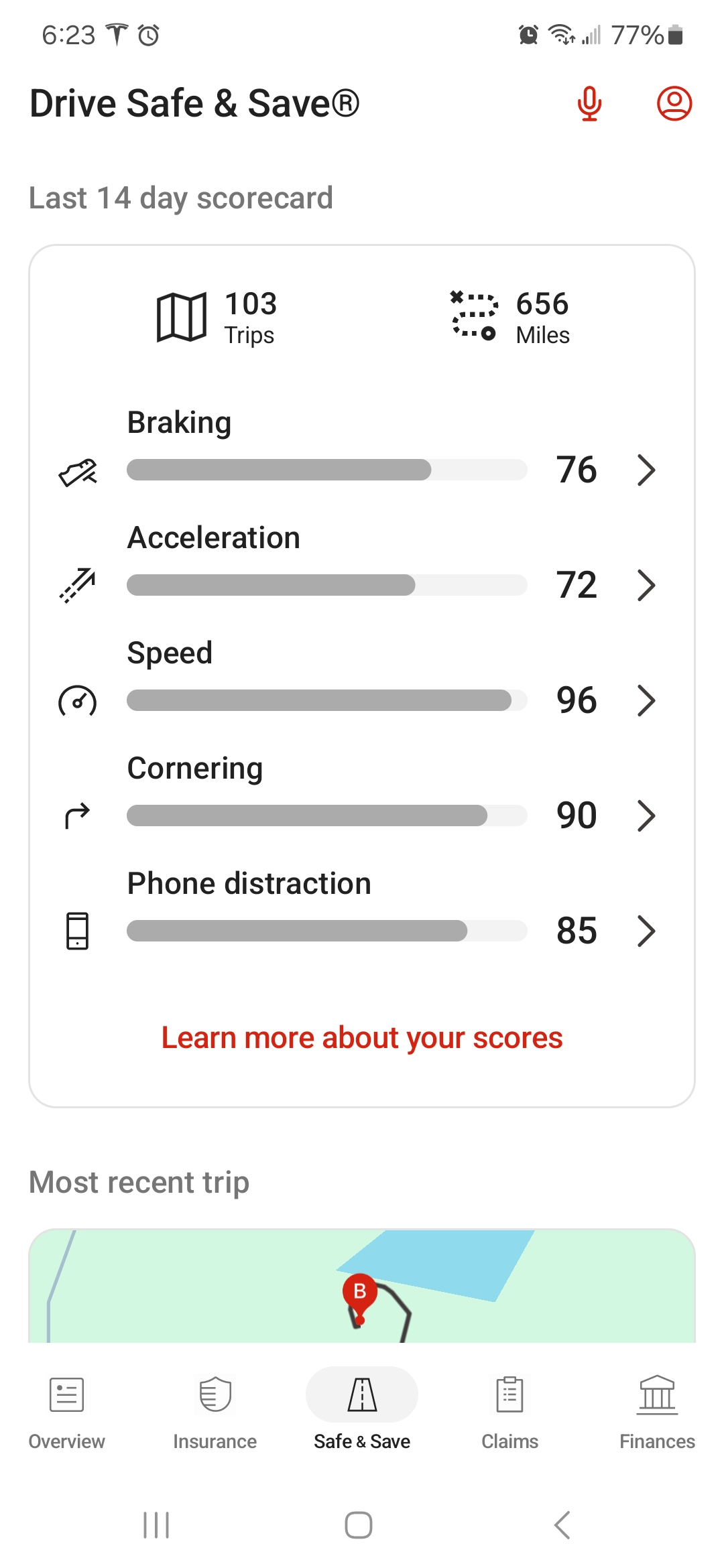}
            \label{statefarm-summary}
            \end{subfigure}
            \hfill
            \begin{subfigure}[t]{0.49\textwidth}
            \centering
            \caption{Detailed trip events}
            \includegraphics[width=0.7\textwidth]{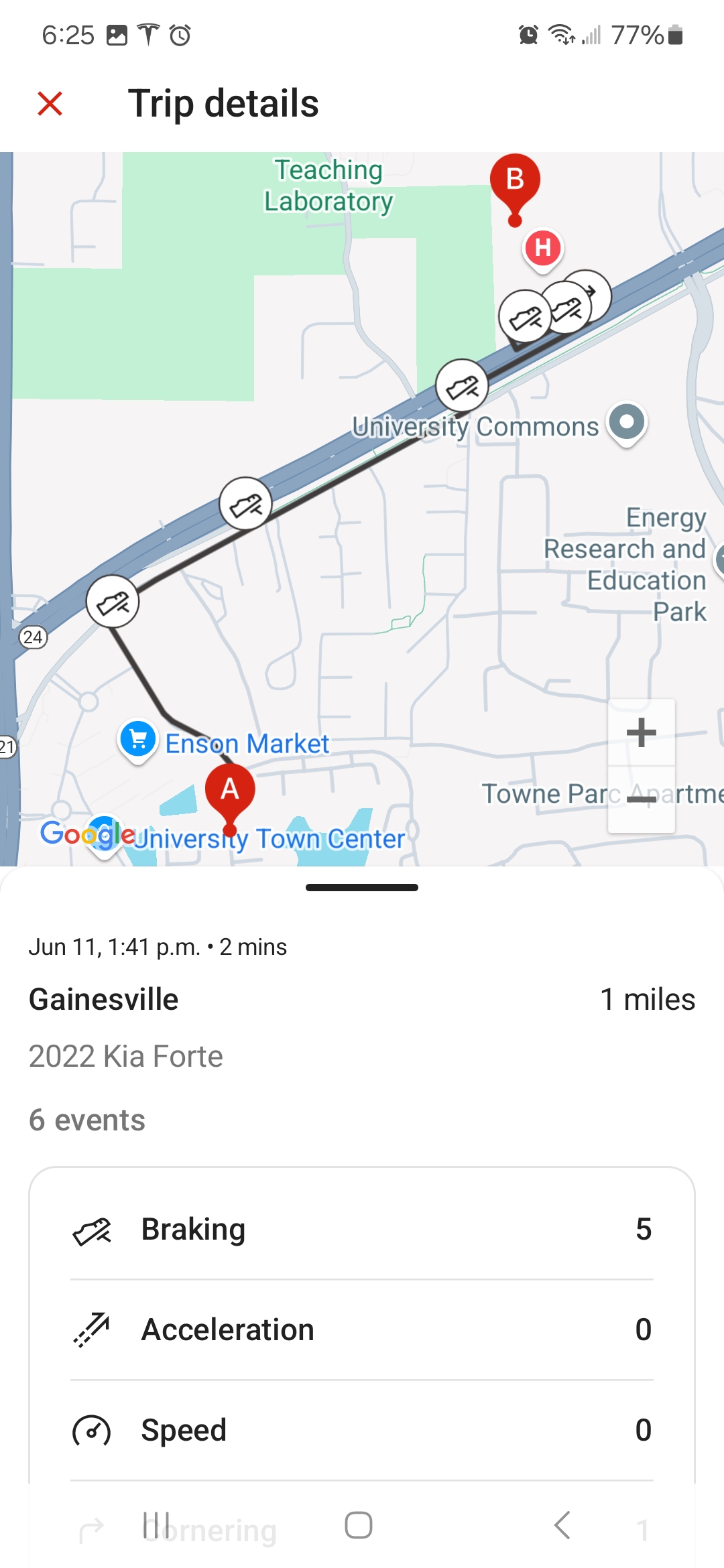}
            \label{statefarm-trip-details}
            \end{subfigure}
        \label{statefarm}
    \end{figure}

    Insurance companies increasingly use driving behavior monitoring to assess accident risk and adjust premiums accordingly. State Farm's Drive Safe \& Save program\footnote{\url{https://www.statefarm.com/insurance/auto/discounts/drive-safe-save}}, for example, offers a discount on insurance premiums to customers who install a monitoring app that tracks braking, acceleration, speeding, and phone distraction in real-time. Figure \ref{statefarm} shows the State Farm monitoring app interface, which provides both overall safety scores and detailed trip-level event tracking.
    
    KCID enables research into CAN bus-based driver profiling for insurance and safety assessments. Key behavioral indicators include:
    
    \begin{itemize}
        \item \textbf{Acceleration:} Accelerator pedal position, engine RPMs, and related signals reveal the frequency and intensity of rapid acceleration events.
        \item \textbf{Braking:} Brake pedal position, speed changes, and braking-related signals distinguish hard braking from gradual deceleration patterns.
        \item \textbf{Turning:} Steering wheel angle combined with vehicle speed indicates whether drivers take corners at appropriate speeds.
        \item \textbf{Signal usage:} Steering wheel angle reveals turning maneuvers. While some angle changes reflect winding roads rather than actual turns, a consistent lack of turn signal activation indicates unsafe behavior.
        \item \textbf{Seatbelt usage:} Many vehicles transmit door and seatbelt status information via CAN, enabling detection of seatbelt non-compliance. Drivers who do not use seatbelts face substantially higher risk of serious injury in crashes.
    \end{itemize}
    
    Machine learning models can classify drivers into risk categories (e.g., very safe, safe, average, aggressive, very aggressive) based on these behavioral patterns extracted from raw CAN data. This approach offers advantages over smartphone-based monitoring by directly accessing vehicle systems rather than inferring behavior from GPS coordinates and phone sensors (gyroscopes and accelerometers). However, researchers should note that true ``unsafe'' driving data collected in live traffic conditions presents ethical and safety challenges. Controlled simulation of unsafe behaviors at a closed-course test facility would be a safer alternative, albeit with potential fidelity limitations.

    In addition to helping insurance companies estimate risk, driver profiling can also help drivers understand their unsafe driving habits in order to become better, safer drivers.

\subsection{Mechanical Anomaly Detection}

    CAN traffic data can facilitate early detection of mechanical problems before traditional warning systems activate. To support research in this domain, we systematically collected raw CAN data under both normal and low tire pressure conditions.
    
    This is one of KCID's specialized experiments (see Section \ref{specialized-experiments}), which we refer to as the ``Tire Pressure Experiment'' in this paper and in the dataset. We have not yet analyzed this data ourselves, but we are providing it to researchers and practitioners who may be interested in investigating whether tire pressure anomalies can be detected through secondary CAN signals beyond direct tire pressure monitoring system (TPMS) warnings.
    
    Potential applications of mechanical anomaly detection include:
    
    \begin{itemize}
        \item \textbf{Tire pressure anomalies:} Detecting low tire pressure through differential wheel speeds or traction control system activation patterns, particularly valuable for vehicles lacking TPMS.
        
        \item \textbf{Sensor failure detection:} Identifying malfunctioning sensors by detecting inconsistencies between related measurements (e.g., mismatches between fuel injection rate and engine RPM).
        
        \item \textbf{Predictive maintenance:} Predicting maintenance needs based on usage patterns rather than fixed mileage intervals. Aggressive driving behaviors (e.g., hard braking, rapid acceleration) accelerate component wear. Driver profile-based wear multipliers (e.g., 1.25x for aggressive drivers, 0.75x for cautious drivers) could improve maintenance scheduling accuracy.
        
        \item \textbf{Fluid system monitoring:} Detecting coolant leaks or other fluid system problems through abnormal temperature patterns before engine overheating occurs.
    \end{itemize}
    
    These applications require vehicle-specific reverse engineering to identify relevant CAN signals. To address this challenge, researchers can leverage automated and semi-automated CAN reverse engineering methods \cite{extracting-signals,can-reverse-engineering,recan-dataset,classification-ids}. Additionally, modified intrusion detection systems could detect anomalies without full signal interpretation.

\subsection{Additional Applications}

    Beyond the primary applications detailed above, KCID supports research in several other domains:
    
    \begin{itemize}
        \item \textbf{Young driver monitoring:} Comparing teenage driver behavior against safe and unsafe driving baselines to provide feedback and identify concerning patterns. This application complements parental oversight and driver education programs.
        
        \item \textbf{Fleet management:} Monitoring driving behavior across company vehicles to identify training needs, reduce accident risk, and optimize vehicle maintenance scheduling.
        
        \item \textbf{Rental vehicle management:} Assessing how drivers treat rental vehicles to adjust pricing or identify excessive wear and tear.
        
        \item \textbf{Restricted driver enforcement:} Implementing geographical or speed restrictions for designated drivers (e.g., teenage drivers limited to specific areas or maximum speeds).
        
        \item \textbf{Reckless driving intervention:} Detecting patterns consistent with reckless driving and providing warnings or implementing vehicle slowdown to prevent dangerous behavior and potential legal consequences. This application is particularly relevant in jurisdictions where vehicle owners face severe penalties (including vehicle seizure) for reckless driving by other operators.
    
        \item \textbf{Impaired driving detection:} While breath alcohol ignition interlock devices (breathalyzers that prevent vehicle operation until the driver provides an alcohol-free breath sample) are widely deployed, these systems can be circumvented if another individual provides the breath sample. CAN bus-based behavioral monitoring offers a complementary approach by comparing real-time driving patterns against established impaired driver profiles. Detection of impaired driving patterns could trigger vehicle warnings or disablement, potentially preventing accidents before impaired drivers endanger others.
    \end{itemize}
    
    These applications demonstrate KCID's versatility for automotive behavior analysis and safety research beyond its primary focus on driver authentication.

\section{Anti-Theft System Framework and Proof of Concept} \label{anti-theft-framework-and-proof-of-concept}

We present a driver authentication anti-theft framework and demonstrate its feasibility through a proof-of-concept prototype. Using a Raspberry Pi 4 with a Sense HAT, we implemented the framework's core functionality and validated it through live vehicle testing. This section describes the proposed framework architecture and presents our proof-of-concept prototype, which can be upgraded to a full working prototype with the addition of driver authentication algorithms.

\subsection{Framework Description}

    The proposed driver authentication anti-theft system uses behavioral biometrics derived from raw CAN bus traffic to distinguish authorized from unauthorized vehicle operators. The system continuously authenticates drivers during vehicle operation, providing a defense-in-depth approach that complements traditional immobilizers. While our focus is raw CAN bus data, the framework also supports decoded OBD-II data.
    
    The system applies machine learning techniques (Section \ref{machine-learning}) to extract behavioral signatures from CAN message patterns. Feature engineering transforms raw CAN data---timestamps, arbitration IDs, and data fields---into discriminative features through temporal analysis and pattern recognition.
    
    Two main approaches enable authentication decisions. Supervised learning methods use labeled training data where some drivers are designated as authorized and others as unauthorized, enabling traditional classification algorithms (decision trees, random forests, support vector machines) to learn to distinguish between the two groups. Unsupervised learning methods rely exclusively on authorized driver data, establishing behavioral baselines through clustering (e.g., $k$-means), autoencoders, etc. that detect unauthorized access through deviation from established patterns. The unsupervised approach addresses the practical challenge that training data from actual vehicle thieves is unavailable.
    
    The framework supports role-based access control with multiple authorization levels. Standard authorized drivers receive unrestricted access, while restricted users (e.g., young drivers) may face geographical boundaries, speed limitations, or time-of-day restrictions. These graduated permissions enable applications beyond theft prevention, including parental oversight and fleet management.

\subsection{Authentication Workflow}

    \begin{diag}[htb!]
        \centering
        \includegraphics[width=0.95\textwidth]{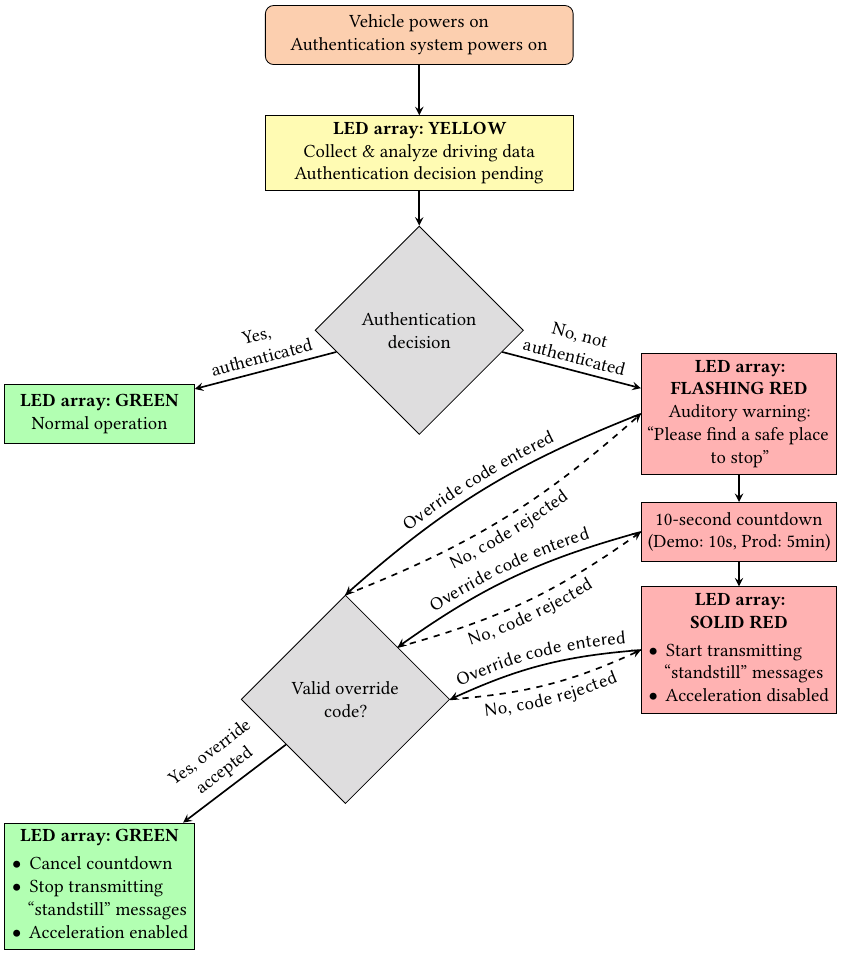}
        \caption{Workflow for the proposed driver authentication anti-theft system. The system continuously monitors driving behavior, providing visual feedback through LED indicators and implementing graduated intervention stages for unauthorized access attempts.}
        \label{auth_workflow}
    \end{diag}

    Upon detecting an unauthorized driver, the system initiates a multi-stage intervention protocol:
    
    \begin{enumerate}
        \item \textbf{Initial Warning:} Visual and auditory alerts notify the driver of authentication failure and request override code verification.
        \item \textbf{Grace Period:} A configurable time window (e.g., 5 minutes) allows the driver to safely stop and enter an owner-provided override code.
        \item \textbf{Acceleration Disable:} Without successful authentication, the system transmits CAN messages that disable acceleration while maintaining braking and steering, forcing the vehicle to coast to a safe stop.
        \item \textbf{Override Mechanism:} Legitimate users can obtain time-limited override codes from vehicle owners, enabling temporary authorization for borrowed vehicles.
    \end{enumerate}

    Diagram \ref{auth_workflow} illustrates this workflow as implemented in our prototype.

\subsection{Disabling Acceleration}

    Our prototype disables acceleration through targeted CAN bus message injection. We identified several CAN frames capable of deceiving the 2011 Chevrolet Traverse's powertrain control systems, preventing acceleration while maintaining essential safety functions (steering, braking).
    
    \textbf{Experimental Methodology.} We tested multiple CAN frames, injecting them repeatedly with brief inter-message delays. Effectiveness varied based on data field content, transmission timing, and vehicle speed. Key findings included:
    
    \begin{itemize}
        \item Disabling acceleration requires brief periods without accelerator input, naturally occurring at traffic signals and intersections
        \item Disabling acceleration causes engine revving without vehicle acceleration
        \item RPM gauge oscillates between injected values (typically zero) and actual engine speed
        \item Some ancillary systems (e.g., air conditioning) may be affected
        \item Brief lag exists between command initiation and system response
    \end{itemize}
    
    \textbf{Effective Command.} We identified a reliable acceleration disable command that proved effective across all tested speeds (up to 35 mph, the maximum safely testable on our route):
    
    \texttt{while true; do cansend can0 0C9\#0000000000001800; sleep 0.001; done}
    
    This command successfully inhibited acceleration at very low speeds (<20 mph), low speeds (20-30 mph), and moderate speeds (30-35 mph), suggesting effectiveness across the vehicle's full operational range. Alternative commands showed similar effectiveness, though testing focused on one command for the proof-of-concept implementation.
    
    \textbf{Safety Considerations.} The acceleration disable mechanism maintains critical safety systems---steering control is unaffected, braking remains fully functional, and the vehicle can coast safely and \textit{gradually} to a stop. This design prioritizes occupant and traffic safety while achieving the anti-theft objective.

\subsection{Proof-of-Concept Prototype}

    \begin{figure*}[htb!]
        \centering
        \includegraphics[width=0.7\textwidth]{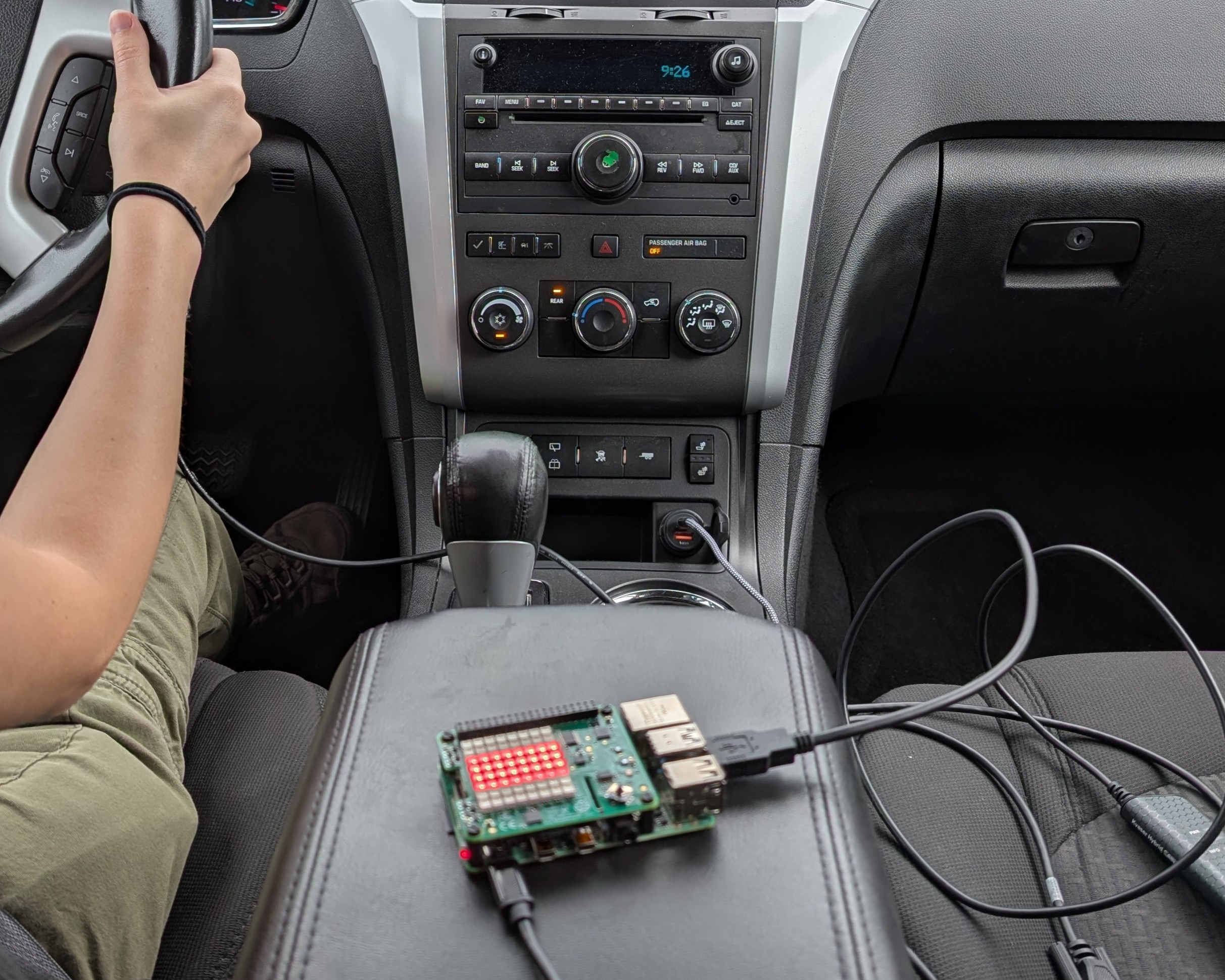}
        \caption{Proof-of-concept prototype implemented on a Raspberry Pi 4 with a Sense HAT. The flashing red LED array indicates authentication failure, warning that acceleration will be disabled after the grace period expires.}
        \label{raspberry-pi-prototype}
    \end{figure*}
    
    Figure \ref{raspberry-pi-prototype} shows our prototype implementation using a Raspberry Pi 4 with a Sense HAT. The system provides visual feedback through an LED array and connects to the vehicle's CAN bus for both data collection and command transmission. This proof-of-concept prototype demonstrates the framework's viability and can be upgraded to a full working prototype with the addition of driver authentication algorithms.
    
    \textbf{System States and Visual Feedback:}
    
    \begin{itemize}
        \item \textbf{Yellow LED (Pending):} Collecting initial driving data, authentication decision pending
        \item \textbf{Green LED (Authenticated):} Driver authenticated, normal operation
        \item \textbf{Flashing Red LED (Warning):} Authentication failed, grace period active
        \item \textbf{Solid Red LED (Disabled):} Grace period expired, acceleration disabled
    \end{itemize}
    
    \textbf{Demonstration Configuration.} For practical demonstration purposes, the prototype provides a 10-second grace period instead of a much longer production setting (e.g., 5 minutes). Joystick inputs simulate authentication decisions (up: \textit{authentication successful} down: \textit{authentication failed}) and override code entry---functions that would be fully implemented in production. While our LED visual warnings worked successfully, auditory warnings did not work without HDMI output and were not implemented. A production system would include warnings such as: ``You are not authenticated. Acceleration will be disabled in five minutes. Please find a safe place to stop and request an override code from the owner of the vehicle.''

    Testing was conducted using the 2011 Chevrolet Traverse---one of the primary vehicles in KCID---on low-traffic public roads. To ensure traffic safety during deceleration tests, we monitored approaching vehicles and temporarily suspended experiments when necessary to avoid impeding traffic flow, resuming testing after allowing vehicles to pass.
    
    Video demonstrations are available on YouTube\footnote{\url{https://youtu.be/VyQa07jvr5Q}} and in Google Drive\footnote{\url{https://drive.google.com/file/d/1f-OlWwrw1-Yuu0FiwBrT4c8GrfQLv1bF/view?usp=sharing}}.
    
\section{Limitations and Future Work} \label{limitations-future-work}

In this section, we examine the limitations inherent in the Kidmose CANid Dataset and proof-of-concept prototype, then we outline promising directions for future research in the domain of driver authentication.

\subsection{Dataset Limitations}

    We designed KCID to address specific gaps in existing driver fingerprinting datasets, prioritizing data types and collection methodologies underrepresented in prior work. This strategic focus means that KCID exhibits different strengths and limitations compared to existing datasets---where other datasets provide extensive coverage, KCID may offer less, and vice versa. Beyond these deliberate design trade-offs, KCID faces practical constraints common to automotive behavioral biometrics research: collecting large-scale driving data representative of diverse demographics, vehicle types, and geographic regions requires substantial time, resources, and logistical coordination.
    
    \textbf{Experimental Design.} KCID prioritizes realistic daily driving data over fixed-route experiments, addressing a critical limitation of existing datasets (Section \ref{research-gaps-and-dataset-limitations}). However, this emphasis results in less controlled route data than datasets like the HCRL Driving Dataset \cite{hcrl-driving-dataset,hcrl-driving-dataset-DATA} or the This Car is Mine! Dataset \cite{this-car-is-mine-dataset,this-car-is-mine-dataset-escar,this-car-is-mine-dataset-DATA}. Similarly, decoded OBD-II data exists for only one driver, as our focus on raw CAN data addresses the predominance of OBD-based datasets in existing literature.
    
    \textbf{Demographic Distribution.} KCID exhibits uneven demographic representation, with more male drivers than female drivers and limited representation in the ``over 55 years of age'' category. The small female cohort (five drivers) precluded age-based subdivision; one or more individuals would be uniquely identified if divided into the same three age categories as the male drivers. While female drivers spanned all three age categories, more balanced representation would strengthen the dataset's utility for evaluating driver authentication performance across diverse user populations.
    
    \textbf{Vehicle Diversity.} The dataset encompasses four vehicles from different manufacturers (Chevrolet, Ford, Subaru, Honda), representing useful diversity but far from comprehensive coverage of automotive makes, models, and model years. Body style variation is limited (three crossover SUVs, one station wagon). Other manufacturers' CAN bus implementations may differ substantially from those represented in KCID. The 2022 Honda CR-V's limited CAN traffic (six arbitration IDs) demonstrates how dramatically vehicles can differ in data availability.
    
    \textbf{Geographic and Cultural Scope.} Data collection occurred in only two countries (United States and Denmark), limiting representation of global traffic laws and driving norms. Cultural factors influence driving behavior, and broader international representation would enhance the dataset's utility.

\subsection{Prototype Limitations}

    Our proof-of-concept prototype demonstrates the feasibility of driver authentication anti-theft systems, but we will need to further develop our proof-of-concept prototype to turn it into a full working prototype. In particular, we will need to implement the driver authentication algorithms, a critical component of a functional driver authentication anti-theft system.
    
    \textbf{Cross-Platform Compatibility.} Even a full working prototype will be several steps short of an anti-theft system suitable for production and real-world deployment. Our proof-of-concept prototype was developed and tested exclusively on the 2011 Chevrolet Traverse and is likely only compatible with similar Chevrolet vehicles. To adapt the prototype to other makes, models, and model years, we will need to reverse engineer their CAN buses. From there, we will need to determine how to disable their accelerators without damaging the vehicles or compromising safety-critical systems.
    
    \textbf{Production Deployment.} Furthermore, a production version must be physically inaccessible to thieves and resistant to tampering. Our current implementation simply connects to the CAN bus through the readily accessible OBD-II port and would be trivial for a thief to disable. Ideally, the system would be equipped with robust tamper-protection features and installed deep within the vehicle, somewhere that is neither quick nor easy to access.

\subsection{Future Research Directions}

    In this section, we discuss future research directions including dataset expansion, signal decoding, driver authentication algorithm development, and investigation of real-world deployment challenges.
    
    \textbf{Dataset Expansion.} One of our priorities is to recruit more female drivers across all age groups to enable age-stratified analysis. In addition, we want to recruit more participants in the ``over 55 years of age'' category for both sexes to strengthen the statistical power for this demographic. Broader vehicle representation, encompassing different body styles (coupes, sedans, minivans, light trucks), additional manufacturers, and newer model years with advanced driver assistance systems, would improve generalizability. Similarly, expanded international data collection encompassing diverse traffic laws and driving norms would enhance cross-cultural applicability. Finally, controlled unsafe driving data collection on closed-course test tracks would enable research on distinguishing safe versus unsafe driving behaviors without exposing participants or the public to risk.
    
    \textbf{Driver Authentication Algorithms.} We plan to extend our proof-of-concept prototype into a fully functional driver authentication system. This will require implementing driver authentication algorithms, likely employing several of the techniques we described in Section \ref{machine-learning}. Our proof-of-concept prototype, implemented on a Raspberry Pi 4 with a Sense HAT, provides a foundation for algorithm development and evaluation. We will prioritize algorithms that balance authentication accuracy with computational efficiency suitable for rugged, reliable automotive embedded systems.
    
    \textbf{Signal decoding.} Community-driven efforts like openDBC\footnote{\url{https://github.com/commaai/opendbc}} facilitate CAN signal interpretation across vehicle models. Analyzing the CAN traffic data of KCID's four vehicles and contributing signal mappings to open-source repositories would enhance the dataset's immediate utility while benefiting the broader automotive cybersecurity research community.
    
    \textbf{Real-World Deployment Challenges.} Practical deployment of driver authentication anti-theft systems requires research across multiple domains:
    
    \begin{itemize}
        \item \textbf{Hardware Security:} Production anti-theft systems will depend on the development of tamper-resistant driver authentication system designs as well as identification of secure vehicle installation locations that prevent physical bypass.
        
        \item \textbf{Regulatory and Legal Frameworks:} Liability issues and regulatory compliance requirements should be investigated, given that the proposed driver authentication anti-theft system will actively intervene in vehicle control.
        
        \item \textbf{Human Factors and Usability:} To ensure that driver authentication systems are not turned off due to false alarms, user acceptance studies should be conducted. These studies should specifically examine driver responses to authentication failures, override mechanisms, and intervention protocols. Research should balance security objectives against usability requirements and investigate interface designs that minimize false rejections while maintaining robust security.
        
        \item \textbf{Safety and Reliability:} Failure mode and effects analysis (FMEA) can help identify potential system failure scenarios and facilitate the development of fail-safe mechanisms ensuring that driver authentication system malfunctions do not create safety hazards or prevent legitimate vehicle operation during emergencies.
    \end{itemize}

\section{Conclusion} \label{conclusion}

This paper makes three key contributions to driver authentication research. First, we provide a comprehensive review of existing open-access driver fingerprinting datasets, identifying critical limitations that constrain the development of robust driver authentication systems. Second, we introduce the Kidmose CANid Dataset (KCID), which addresses these limitations by providing raw CAN bus data from 16 drivers across four vehicles under realistic driving conditions. Third, we present a driver authentication anti-theft framework with a proof-of-concept implementation, demonstrating practical feasibility through live road trials with an unaltered passenger vehicle.

KCID represents a substantial advancement in publicly available driver fingerprinting datasets. Unlike existing datasets that rely on decoded diagnostic data collected at low sampling rates over artificial fixed routes, KCID provides raw CAN bus traffic captured during realistic driving conditions. The inclusion of demographic information enables researchers to evaluate whether driver authentication systems can distinguish between drivers with similar profiles---a critical capability absent from existing datasets. The dataset also supports diverse applications beyond authentication, including insurance risk assessment, mechanical anomaly detection, and impaired driving detection.

While the dataset would benefit from additional drivers and vehicles to enable more granular demographic analysis, KCID addresses the most critical methodological gaps in existing resources. Our proof-of-concept implementation demonstrates that driver authentication systems can be deployed on standard hardware in unaltered vehicles, establishing practical feasibility for real-world adoption. KCID provides researchers with both the data and methodological foundation necessary to advance driver authentication research and develop deployable systems that enhance automotive cybersecurity through behavioral biometrics.

\section{Acknowledgments} \label{acknowledgments}

We thank our volunteer drivers, especially Lars Brasen, Marius Brasen, Christian Lampe, Julie Lampe, Ryan Lampe, Alan \& Jody Olsen, Brian \& Deb Olsen, Chris Olsen, Eric Olsen, and Wayne \& Vickie Olsen for allowing us to collect and publish their driving data during their normal daily driving activities, and for permitting us to use their demographic information to identify the data in our research.

This work was partially supported by Innovation Fund Denmark project CyberQ (IFD project no. 3200-00035B).

\bibliographystyle{unsrt}
\bibliography{main}

\end{document}